\pdfoutput=1

\documentclass[11pt,twoside,a4paper,cmspaper,final,collab]{cms-tdr}
\def\svnVersion{302626}\def\cmsCernNo{2013-037}\def\cmsCernDate{\today}\def\cmsMessage{Published in Physical Review D as \href{http://dx.doi.org/10.1103/PhysRevD.92.032008}{\doi{10.1103/PhysRevD.92.032008.}}}

\begin{document}\cmsNoteHeader{HIG-14-004}

\hyphenation{had-ron-i-za-tion}
\hyphenation{cal-or-i-me-ter}
\hyphenation{de-vices}
\RCS$HeadURL: svn+ssh://svn.cern.ch/reps/tdr2/papers/HIG-14-004/trunk/HIG-14-004.tex $
\RCS$Id: HIG-14-004.tex 302320 2015-09-02 14:52:21Z salderwe $
\newlength\cmsFigWidth
\ifthenelse{\boolean{cms@external}}{\setlength\cmsFigWidth{0.98\columnwidth}}{\setlength\cmsFigWidth{0.6\textwidth}}
\ifthenelse{\boolean{cms@external}}{\providecommand{\cmsLeft}{top}}{\providecommand{\cmsLeft}{upper left}}
\ifthenelse{\boolean{cms@external}}{\providecommand{\cmsRight}{bottom}}{\providecommand{\cmsRight}{right}}
\ifthenelse{\boolean{cms@external}}{\providecommand{\CL}{CL\xspace}}{\providecommand{\CL}{CL\xspace}}
\providecommand{\CLs}{\ensuremath{\mathrm{CL}_\mathrm{s}}\xspace}
\cmsNoteHeader{HIG-14-004}

\title{Search for the standard model Higgs boson produced through vector boson fusion and decaying to \texorpdfstring{\bbbar}{b bbar}}

\date{\today}

\abstract{
A first search is reported for a standard model Higgs boson (\PH) that is produced through vector boson fusion and decays to a bottom-quark pair. Two data samples, corresponding to integrated luminosities of 19.8\fbinv and 18.3\fbinv of proton-proton collisions at $\sqrt{s}=8$\TeV were selected for this channel at the CERN LHC. The observed significance in these data samples for a
$\PH\to\bbbar$ signal at a mass of 125\GeV is 2.2 standard deviations, while the expected significance is 0.8 standard deviations. The fitted signal strength $\mu=\sigma/\sigma_\mathrm{SM}=2.8^{+1.6}_{-1.4}$. The combination of this result with other CMS searches for the Higgs boson decaying to a \PQb-quark pair, yields a signal strength of $1.0\pm 0.4$, corresponding to a signal significance of 2.6 standard deviations for a Higgs boson mass of 125\GeV.
}

\hypersetup{%
pdfauthor={CMS Collaboration},%
pdftitle={Search for the standard model Higgs boson produced through vector boson fusion and decaying to b bbar},%
pdfsubject={CMS},%
pdfkeywords={CMS, physics, Higgs, VBF, Hbb}}

\maketitle

\section{Introduction}

In the standard model (SM) \cite{SM3,Glashow:1961tr,Weinberg:1967tq}, the electroweak symmetry breaking is achieved by a mechanism~\cite{PhysRevLett.13.321,PhysRevLett.13.508,PhysRevLett.13.585} that provides mass to the electroweak gauge bosons, while leaving the photon massless. The mechanism predicts the existence of a scalar Higgs boson (\textit{H}), and its observation was one of the main goals of the CERN LHC program. A boson with mass near 125\GeV was recently discovered by both the ATLAS~\cite{:2012gk} and CMS~\cite{:2012gu,Chatrchyan:2013lba} collaborations, with properties that are compatible with those of a SM Higgs boson~\cite{Aad:2013wqa,Khachatryan:2014jba}.

At the LHC, a SM Higgs boson can be produced through a variety of mechanisms. The expected production cross sections~\cite{Dittmaier:2012vm} as a function of the Higgs boson mass are such that, in the mass range considered in this study, the vector boson fusion (VBF) process $\Pp\Pp\to\Pq\Pq\PH$ has the second largest production cross section following gluon fusion (GF). Furthermore, for a SM Higgs boson with a mass $m_{\PH}\lesssim 135\GeV$, the expected dominant decay mode is to a \PQb-quark pair ($\cPqb\cPaqb$).

Thus far, the search for $\PH\to\cPqb\cPaqb$ has been carried out in the associated production process involving a $\PW$ or a $\Z$ boson ($V\PH$ production) at the Tevatron~\cite{TevatronVH} and at the LHC~\cite{CMS-VHbb,ATLAS-VHbb}, as well as in association with a top quark pair at the LHC~\cite{CMS-ttHbb,CMS-ttH,Aad:2015gra}, without reaching the necessary sensitivity to observe the Higgs boson in this decay channel. It is therefore important to exploit other production modes, such as VBF, to provide in the $\cPqb\cPaqb$ decay channel further information on the nature and properties of the Higgs boson.

The prominent feature of the VBF process $\Pq\Pq\PH\to\Pq\Pq\cPqb\cPaqb$ is the presence of four energetic jets in the final state. Two jets are expected to originate from a light-quark pair ($\PQu$ or $\PQd$), which are typically two valence quarks from each of the colliding protons scattered away from the beam line in the VBF process. These ``VBF-tagging'' jets are expected to be roughly in the forward and backward directions relative to the beam direction. Two additional jets are expected from the Higgs boson decay to a $\cPqb\cPaqb$ pair in more central regions of the detector. Another important property of the signal events is that, being produced through an electroweak process, no quantum chromodynamics (QCD) color is exchanged at leading order in the production. As a result, in the most probable color evolution of these events, the VBF-tagging jets connect to the proton remnants in the forward and backward beam line directions, while the two \PQb-quark jets connect to each other as decay products of the color neutral Higgs boson. Consequently very little additional QCD radiation and hadronic activity is expected in the space outside the color-connected regions, in particular in the whole rapidity interval (rapidity gap) between the two VBF-tagging jets, with the exception of the Higgs boson decay products.

The dominant background to this search is from QCD production of multijet events. Other backgrounds arise from: (i) hadronic decays of $\Z$ or $\PW$ bosons produced in association with additional jets, (ii) hadronic decays of top quark pairs, and (iii) hadronic decays of singly produced top quarks. The contribution of the Higgs boson in GF processes with two or more associated jets is included in the expected signal yield.

The search is performed on selected four-jet events that are characterized by the response of a multivariate discriminant trained
to separate signal events from background without making use of kinematic information on the two \PQb-jet candidates. Subsequently, the invariant mass distribution of two $\PQb$ jets is analyzed in each category in the search for a signal ``bump'' above the smooth contribution from the SM background. This is the first search of this kind, and the only search for the SM Higgs boson in all-jet final states at the LHC.
A search for a SM Higgs boson in the all-hadronic final state has been previously reported
by the CDF experiment~\cite{Aaltonen:2012ji}.

This paper is organized as follows: Section~\ref{sec:det} highlights the features of the CMS detector needed to perform this analysis. Section~\ref{sec:sim} details the production of simulated samples used to study the signal and main backgrounds,
and Section~\ref{sec:trig} presents the employed triggers. Event reconstruction and selection are described in Sections~\ref{sec:reco} and~\ref{sec:sel}, respectively. The unique features of the analysis are discussed in Section~\ref{sec:special}, which includes the improvement of the resolution in jet transverse momentum (\pt) by regression techniques, discrimination between quark- and gluon-originated jets, and soft QCD activity. An important validation of the analysis strategy is the observation of the $\Z\to\cPqb\cPaqb$ decay, which is presented in Section~\ref{sec:z}. The search for a SM Higgs boson is discussed in Section~\ref{sec:higgs} and the associated systematic uncertainties are presented in Section~\ref{sec:unc}. The final results are discussed in Section~\ref{sec:res} and combined with previous searches in the same channel in Section~\ref{sec:comb}. We summarize in Section~\ref{sec:sum}.

\section{The CMS detector}\label{sec:det}

The central feature of the CMS apparatus is a superconducting solenoid of 6\unit{m} internal diameter, providing a magnetic field of 3.8\unit{T}. A silicon pixel and strip tracker, a lead tungstate crystal electromagnetic calorimeter, and a brass and scintillator hadron calorimeter are located within the solenoidal field. Muons are measured in gas-ionization detectors embedded in the steel flux-return yoke of the solenoid. Forward calorimetry (pseudorapidity $3<\abs{\eta}<5$) complements the coverage provided by the barrel ($\abs{\eta}<1.3$) and end cap ($1.3<\abs{\eta}<3$) detectors. The first level (L1) of the CMS trigger system, composed of specialized processors, uses information from the calorimeters and muon detectors to select the most interesting events in a time interval of less than 4\mus. The high-level trigger (HLT) processor farm decreases the event rate from about 100\unit{kHz} to less than 1\unit{kHz}, before data storage. A more detailed description of the CMS apparatus and the main kinematic variables used in the analysis can be found in Ref.~\cite{Chatrchyan:2008zzk}.

\section{Simulated samples}\label{sec:sim}

Samples of simulated Monte~Carlo (MC) signal and background events are used to guide the analysis optimization and to estimate signal yields. Several event generators are used to produce the MC events. The samples of VBF and GF signal processes are generated using the next-to-leading order perturbative QCD program \POWHEG 1.0~\cite{Nason:2004rx}, interfaced to \TAUOLA 2.7~\cite{Golonka:2003xt} and \PYTHIA 6.4.26~\cite{Sjostrand:2006za} for the hadronization process and modeling of the underlying event (UE). The most recent \PYTHIA 6 Z2* tune is derived from the Z1 tune~\cite{Field:2010}, which uses the CTEQ5L parton distribution functions (PDF), whereas Z2* adopts CTEQ6L~\cite{Pumplin:2002vw}. The signal samples are generated using only $\PH\to\cPqb\cPaqb$ decays, for five mass hypotheses: $m_{\PH}=115$, 120, 125, 130, and 135\GeV.

Background samples of QCD multijet, \Z+jets, \PW+jets, and $\ttbar$ events are simulated using leading-order (LO) \MADGRAPH 5.1.3.2~\cite{Alwall:2011uj} interfaced with \PYTHIA. The single top quark background samples are produced using \POWHEG, interfaced with \TAUOLA and \PYTHIA. The default set of PDF used with \POWHEG samples is CT10~\cite{Lai:2010vv}, while the LO CTEQ6L1 set~\cite{Pumplin:2002vw} is used for other samples. The production cross sections for \PW+jets and \Z+jets are rescaled to next-to-next-to-leading-order (NNLO) cross sections calculated using the \FEWZ 3.1 program~\cite{Gavin:2010az,Li:2012wna,Gavin:2012sy}. The $\ttbar$ and single top quark samples are also rescaled to their cross sections based on NNLO calculations~\cite{Czakon:2013goa, Kidonakis:2012db}.

To accurately simulate the LHC luminosity conditions during data taking, additional simulated $\Pp\Pp$ interactions overlapping in the same or neighboring bunch crossings of the main interaction, denoted as pileup, are added to the simulated events with a multiplicity distribution that matches the one in the data.

\section{Triggers}\label{sec:trig}

The data used for this analysis were collected using two different trigger strategies that result in two different data samples for analysis.
First, a set of dedicated trigger event selection (paths) was specifically designed and deployed for the VBF $\Pq\Pq\PH\to \Pq\Pq\cPqb\cPaqb$ signal search, both for the L1 trigger and the HLT, and operated during the full 2012 data taking. Then, a more general trigger was employed for the larger part of the 2012 data taking that targeted VBF signatures in general. The first (nominal) set of triggers collected the larger fraction of the signal events, while the second trigger supplemented the search with events that failed the stringent nominal-trigger requirements.
The integrated luminosity collected with the first set of triggers was 19.8\fbinv, while for the second trigger it was 18.3\fbinv.

While the first dedicated trigger paths collected data within the standard CMS streams, the second general-purpose VBF trigger path ran
in parallel with data streams that were reconstructed later, in 2013, during the LHC upgrade.

\subsection{Dedicated signal trigger}

The L1 paths require the presence of at least three jets with $\pt$ above decreasing thresholds $\pt^{(1)},\,\pt^{(2)},\,\pt^{(3)}$ that were adjusted according to the instantaneous luminosity [$\pt^{(1)} = 64$--68\GeV, $\pt^{(2)} = 44$--48\GeV, $\pt^{(3)} = 24$--32\GeV]. Among the three jets, one and only one of the two leading jets [with $\pt > \pt^{(1)},\,\pt^{(2)}$] can be in the forward region with pseudorapidity $2.6 < \abs{\eta}\leq 5.2$, while the other two jets are required to be central ($\abs{\eta} \leq 2.6$).

The HLT paths are seeded by the L1 paths described above, and require the presence of
four jets with \pt above thresholds that were again adjusted according to the
instantaneous luminosity, $\pt > 75$--82, 55--65, 35--48, and 20--35\GeV, respectively.
Two complementary HLT paths have been employed that make use, respectively, of
(i) only calorimeter-based jets (CaloJets) and (ii) particle-flow jets (PFJets, see Section~\ref{sec:reco}).
At least one of the four selected jets must further fulfill minimum \PQb-tagging requirements,
evaluated using HLT regional tracking around the jets, and using
the ``track counting high-efficiency'' (TCHE) or the ``combined secondary vertex''
(CSV) algorithms~\cite{Chatrchyan:2012jua}, alternatively for the first and second paths.
Events are accepted if they satisfy either of the two paths.

Among the four leading jets, the light-quark ($\PQq\PQq$) VBF-tagging jet pair is assigned in one of two ways:
(i) the pair with the smallest HLT \PQb-tagging output values (\PQb-tag-sorted $\PQq\PQq$) or
(ii) the pair with the maximum pseudorapidity difference ($\eta$-sorted $\PQq\PQq$).
Both pairs are required to exceed variable minimum thresholds on $\abs{\Delta\eta_{\PQq\PQq}}$ of 2.2--2.5,
and of 200--240\GeV on the dijet invariant mass $m_{\PQq\PQq}$, depending on the instantaneous luminosity.

To evaluate trigger efficiencies, a prescaled control path is used, requiring one PFJet with $\pt > 80\GeV$. To match the efficiency in data, the simulated trigger efficiency must be corrected with a scale factor of order $0.75$ that is parametrized a function of the highest jet \PQb-tag output in the event and the invariant mass of the quark-jet candidates.
With this procedure the weak dependence of the trigger efficiencies on the invariant mass of
the two $\PQb$ jets is also taken into account.

\subsection{General-purpose VBF trigger}

The L1 paths for the general-purpose VBF trigger require that the scalar \pt sum of the hadronic activity in the event exceeds $175$ or $200$\GeV, depending on the instantaneous luminosity.

The HLT path is seeded by the L1 path described above, and requires the presence of at least two CaloJets with $\pt>35\GeV$. Out of all the possible jet pairs in the event, with one jet lying at positive and the other at negative $\eta$, the pair with the highest invariant mass is selected as the most probable VBF-tagging jet pair. The corresponding invariant mass $m_\mathrm{jj}^\text{trig}$
and absolute pseudorapidity difference $\abs{\Delta\eta_\mathrm{jj}^\text{trig}}$ are required to be larger than $700\GeV$ and $3.5$, respectively.

The efficiency of the general-purpose VBF trigger is measured in a similar way as for the dedicated triggers, using a prescaled path (requiring two PFJets with average $\pt > 80\GeV$). To match the efficiency in data, the simulated trigger efficiency must be corrected with a scale factor of order 0.8 that is expressed as a function of the invariant mass and the pseudorapidity difference of the two offline quark-jet candidates.

\section{Event reconstruction}\label{sec:reco}

The offline analysis uses reconstructed charged-particle tracks and candidates from the particle-flow (PF) algorithm~\cite{CMS-PAS-PFT-09-001,CMS-PAS-PFT-10-001,CMS-PAS-PFT-10-002}. In the PF event reconstruction all stable particles in the event, \ie electrons, muons, photons, and charged and neutral hadrons, are reconstructed as PF candidates using information from all CMS subdetectors to obtain an optimal determination of their direction, energy, and type. The PF candidates are then used to reconstruct the jets and missing transverse energy.

Jets are reconstructed by clustering PF candidates with the anti-\kt algorithm~\cite{Cacciari:2005hq,Cacciari:2008gp} with a distance parameter of 0.5. Reconstructed jets require a small additional energy correction, mostly due to thresholds on reconstructed tracks and clusters in the PF algorithm and various reconstruction inefficiencies~\cite{JES}. Jet identification criteria are also applied to reject misreconstructed jets resulting from detector noise, as well as jets heavily contaminated with pileup energy (clustering of energy deposits not associated with a parton from the primary $\Pp\Pp$ interaction)~\cite{CMS-PAS-JME-13-005}. The efficiency of the jet identification criteria is greater than 99\%, with the rejection of 90\% of background pileup jets with $\pt\simeq 50\GeV$.

The identification of jets that originate from the hadronization of $\PQb$ quarks is done with the CSV $\PQb$ tagger~\cite{Chatrchyan:2012jua}, also implemented for the HLT paths, as described in Section~\ref{sec:trig}. The CSV algorithm combines the information from track impact parameters and secondary vertices identified within a given jet, and provides a continuous discriminator output.

Events are required to have at least four reconstructed jets. All the jets found in an event are ordered according to their \pt, and the four leading ones are considered as the most probable $\PQb$ jet and VBF-tagging jet candidates. The distinction between the two jet types is done by the means of a multivariate discriminant that, in addition to the \PQb-tag values and the \PQb-tag ordering, takes into account the $\eta$ values and the $\eta$ ordering. In the VBF $\PH\to\cPqb\cPaqb$ signal simulation
it is found that the $\PQb$ jets have higher \PQb-tag values and are more central in $\eta$ than the VBF-tagging jets. A boosted decision tree (BDT), implemented with the TMVA package~\cite{Hocker:2007ht}, is trained on simulated signal events using the discriminating variables previously described and its output is used as a \PQb-jet likelihood score; out of the four leading jets the two with the highest score are identified as the $\PQb$ jets, while the other two are identified as the VBF-tagging jets. With the use of the multivariate \PQb-jet assignment the signal efficiency is increased by $\approx$10\% compared to the interpretation based on CSV output only.

\section{Event selection}\label{sec:sel}

The offline event selection is based upon the \PQb-jet and VBF-tagging jet assignment described in Section~\ref{sec:reco}, and is adjusted to the two different trigger sets presented in Section~\ref{sec:trig}. In what follows the selected events are divided into two sets referred to as {\it set A} and {\it set B}. These selections are summarized in Table~\ref{tab:sel}.

\begin{figure}[hbt]
  \centering
    \includegraphics[width=0.48\textwidth]{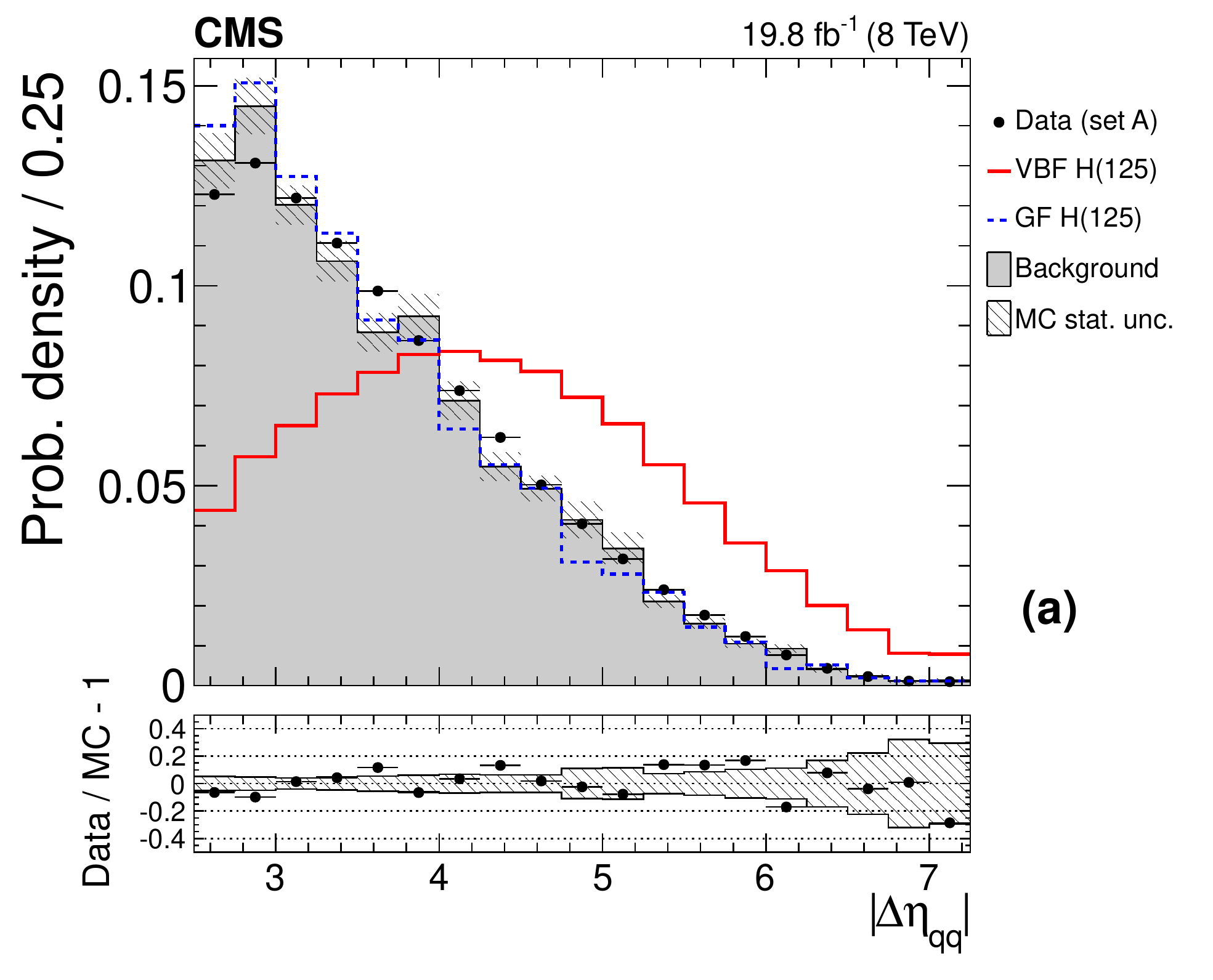}
    \includegraphics[width=0.48\textwidth]{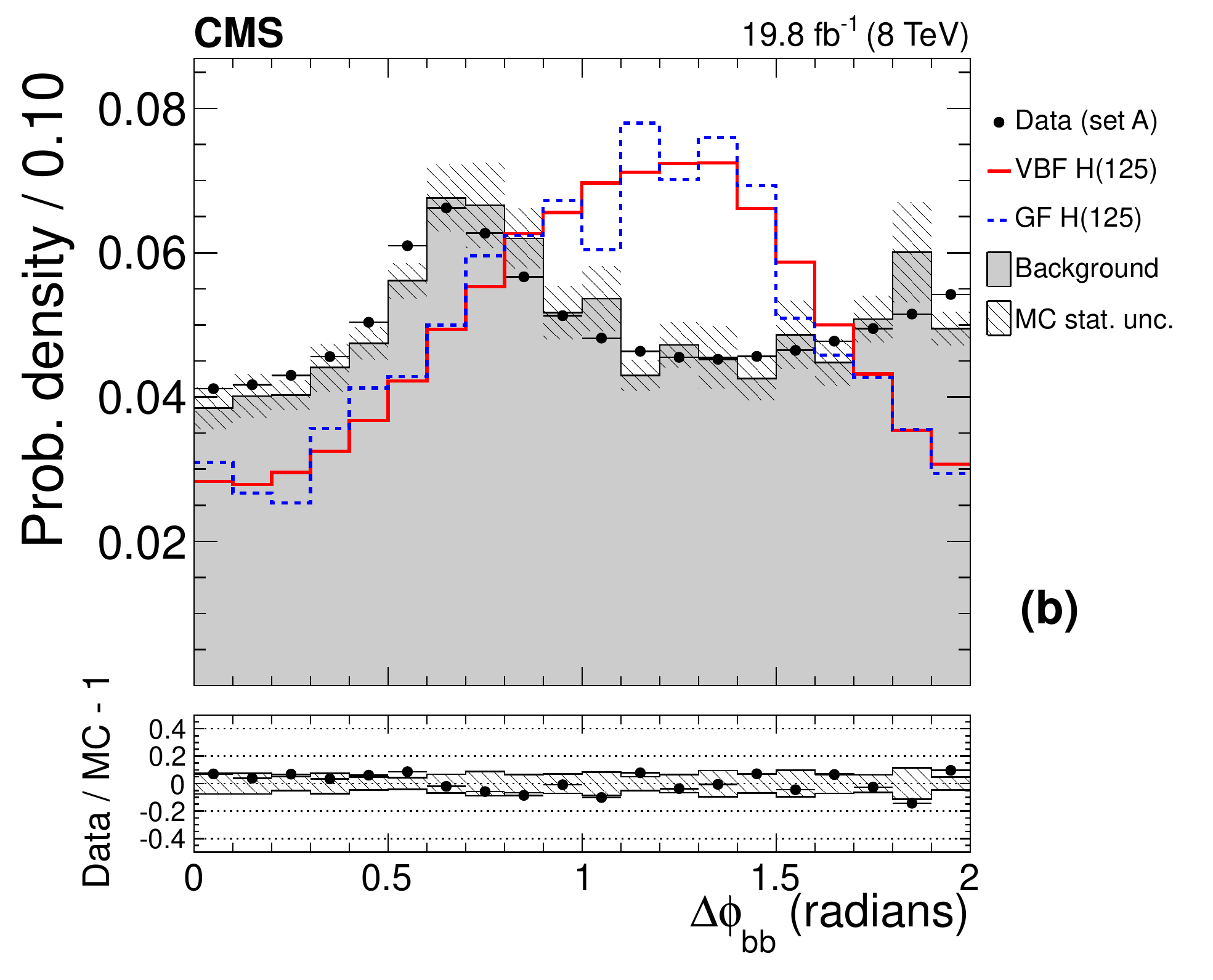}
    \caption{(a) Normalized distribution in absolute pseudorapidity difference between the two VBF-jet candidates $(\abs{\Delta\eta_{\PQq\PQq}})$. (b) Normalized distribution of the azimuthal difference between the two \PQb-jet candidates $(\Delta\phi_{\PQb\PQb})$. The selection corresponds to set A, data are shown by the points, and the sum of all simulated backgrounds is by the filled histograms. The VBF Higgs boson signal is displayed by a solid line, and the GF Higgs boson signal is shown by a dashed line. The panels at the bottom show the fractional difference between data and background simulation, with the shaded band representing the statistical uncertainties in the MC samples.}
    \label{fig:evtSel}

\end{figure}

Events selected in set A are required to have been selected by the dedicated VBF $\Pq\Pq\PH\to \Pq\Pq\cPqb\cPaqb$ trigger and to have at least four PF jets with $\pt^{1,2,3,4}>80,70,50,40\GeV$ and $\abs{\eta}<4.5$. Moreover, at least two of these jets must satisfy the loose CSV working point requirement (CSVL)~\cite{Chatrchyan:2012jua}. The VBF topology is ensured by requiring $m_{\PQq\PQq}>250\GeV$ and $\abs{\Delta\eta_{\PQq\PQq}} > 2.5$, where ${\PQq\PQq}$ denotes the pair of the most probable VBF-tagging jets. Finally, in order to suppress further the background, the azimuthal angle difference $\Delta\phi_{\PQb\PQb}$ between the two \PQb-jet candidates must be less than $2.0$ radians. Figure~\ref{fig:evtSel} shows the normalized distributions of $\abs{\Delta\eta_{\PQq\PQq}}$ (left) and $\Delta\phi_{\PQb\PQb}$ (right) for the sum of all simulated backgrounds, and the VBF and GF Higgs boson production.

Events in set B are first required to not belong to set A (to avoid double counting). Then, they must have passed the generic VBF topological trigger and have at least four PF jets with $\pt>30\GeV$ and $\abs{\eta}<4.5$. In addition, the scalar \pt sum of the two leading jets must be greater than 160\GeV. In order to enrich the sample in $\PQb$ jets, there must be at least one jet satisfying
the medium CSV working point requirement (CSVM)~\cite{Chatrchyan:2012jua} and one jet satisfying the CSVL. The VBF topology is ensured by requiring $m_{\PQq\PQq},\,m_\mathrm{jj}^\text{trig}>700\GeV$, and $\abs{\Delta\eta_{\PQq\PQq}},\,\abs{\Delta\eta_\mathrm{jj}^\text{trig}}>3.5$,
 where ${\PQq\PQq}$ denotes the pair of the most probable VBF jets and $\mathrm{jj}$ denotes the jet pair with the highest invariant mass (as in the trigger logic described in Section~\ref{sec:trig}). Finally the azimuthal angle $\Delta\phi_{\PQb\PQb}$ between the two \PQb-jet candidates must be less than 2.0 radians.

After all the selection requirements, 2.3\% of the simulated VBF signal events (for $m_{\PH}=125\GeV$) end up in set A, and 0.8\% end up in set B. The fraction of events in set A that also satisfy the requirements of set B (except for the set A veto) amounts to 39\%. The set B selection recovers signal events presenting pronounced VBF jets, with lower $\pt$ but larger pseudorapidity opening and invariant mass.

\begin{table*}[htbH]
  \centering
    \topcaption{Summary of selection requirements for the two analyses.}
    \begin{scotch}{lcc}
                        & Set A                  & Set B   \\
      \hline
      \hline
      Trigger           & Dedicated VBF $\Pq\Pq\PH\to \Pq\Pq\cPqb\cPaqb$ & General-purpose VBF trigger \\
      \hline
      \multirow{2}{*}{} & \multirow{2}{*}{}                     &  $\pt^{1,2,3,4}>30\GeV$ \\
        Jets \pt        & $\pt^{1,2,3,4}>80,70,50,40\GeV$  &  \\
                        &                                       &  $\pt^1+\pt^2>160\GeV$ \\
      \hline
      Jets $\abs{\eta}$     & $<$4.5                                & $<$4.5 \\
      \hline
      $\PQb$ tag             & At least 2 CSVL jets                  & At least 1 CSVM and 1 CSVL jets \\
      \hline
      $\Delta\phi_{\PQb\PQb}$ & $<$2.0\unit{radians}                           & $<$2.0\unit{radians} \\
      \hline
      \multirow{2}{*}{} & $m_{\PQq\PQq}>250\GeV$                 & $m_{\PQq\PQq},\,m_\mathrm{jj}^\text{trig}>700\GeV$ \\
        VBF topology    &                                       &  \\
                        & $\abs{\Delta\eta_{\PQq\PQq}} > 2.5$        & $\abs{\Delta\eta_{\PQq\PQq}},\,\abs{\Delta\eta_\mathrm{jj}^\text{trig}}>3.5$ \\
      \hline
      Veto              & None                                  & Events that belong to set A \\
    \end{scotch}
    \label{tab:sel}

\end{table*}

\section{Signal properties}\label{sec:special}

The analysis described in this paper relies on certain characteristic properties of the studied final state, which provide a significant improvement of the overall sensitivity. First, the resolution of the invariant mass of the two $\PQb$ jets is improved by applying multivariate regression techniques. Then, the jet composition properties are used to form a discriminant that separates jets originating from light quarks or gluons. Third, soft QCD activity outside the jets is quantified and used as a discriminant between QCD processes with strong color flow and the VBF signal without color flow.

\subsection{Jet transverse-momentum regression}\label{sec:reg}

The $\cPqb\cPaqb$ mass resolution is improved by using a regression technique similar to those used in the search for a Higgs boson produced in association with a weak vector boson and decaying to $\cPqb\cPaqb$~\cite{CMS-VHbb}. A refined calibration is carried out for individual $\PQb$ jets, beyond the default jet energy corrections, that takes into account the jet composition properties and targets semileptonic $\PQb$ decays that lead to a substantial mismeasurement of the jet \pt due to the presence of an escaping neutrino. For this purpose a regression BDT is trained on simulated signal events with inputs including information about the jet properties and structure. The target of the regression is the \pt of the associated particle-level jet, clustered from all stable particles (with lifetime $c\tau > 1$\unit{cm}). The inputs include:
(i) the jet $\pt$, $\eta$, and mass;
(ii) the jet energy fractions carried by neutral hadrons and photons~\cite{CMS-PAS-PFT-09-001,CMS-PAS-PFT-10-001,CMS-PAS-PFT-10-002};
(iii) the mass and the uncertainty in the decay length associated with the secondary vertex, when present;
(iv)  the event missing transverse energy and its azimuthal direction relative to the jet;
(v) the total number of jet constituents;
(vi) the \pt of the soft-lepton candidate inside the jet, when present, and its \pt component perpendicular to the jet axis;
(vii) the \pt of the leading track in the jet;
and (viii) the average \pt density of the event in $y$--$\phi$ space ({\FASTJET} $\rho$ method~\cite{Cacciari:2007fd}).

\begin{figure}[hbt]
\centering
    \includegraphics[width=\cmsFigWidth]{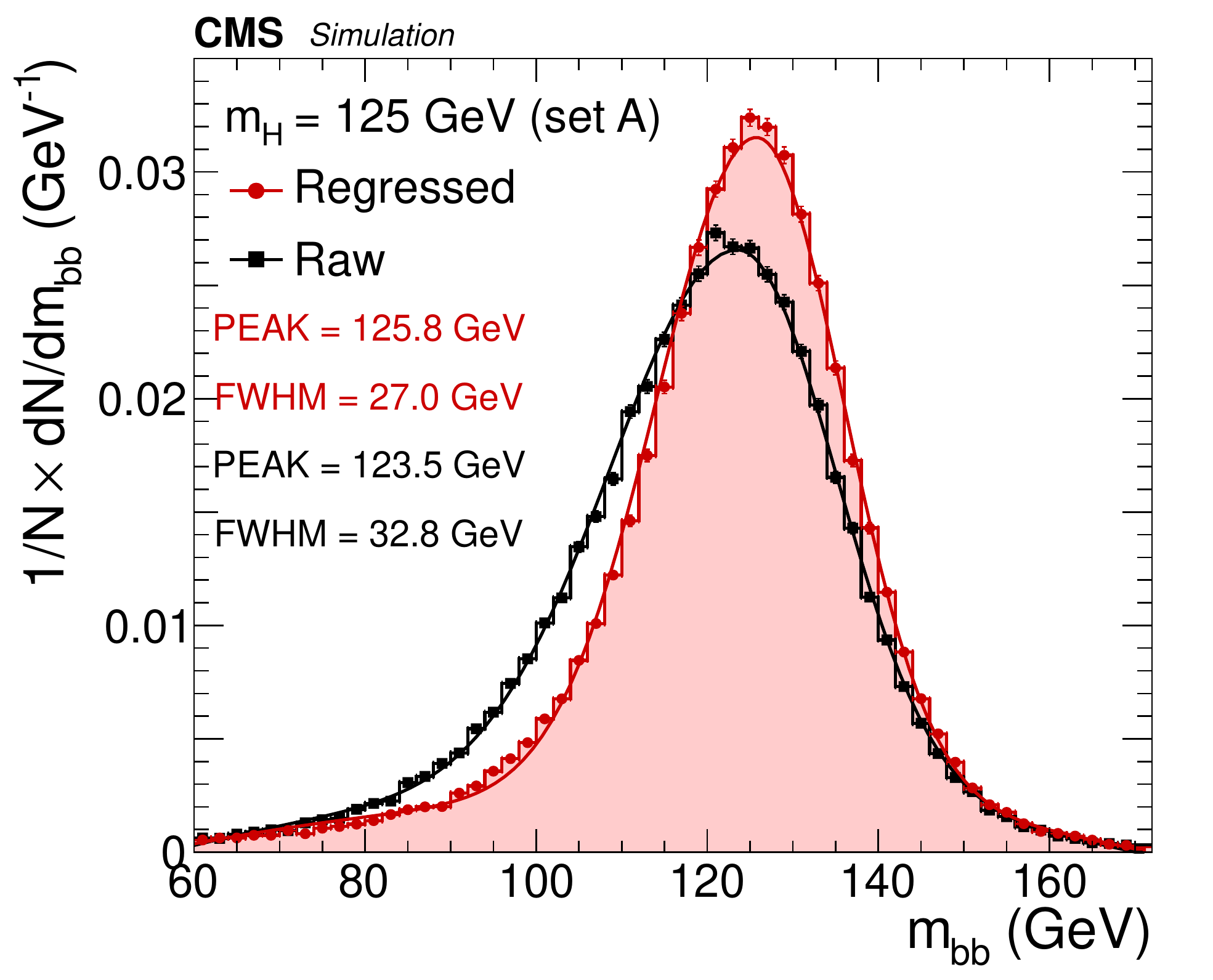}
    \caption{Simulated invariant mass distribution of the two \PQb-jet candidates before and after the jet \pt regression, for VBF signal events. The generated Higgs boson signal mass is 125\GeV and the event selection corresponds to set A. By FWHM we denote the width of the distribution at the middle of its maximum height.}
    \label{fig:mbbSignal}
\end{figure}

\begin{figure}[hbt]
  \centering
    \includegraphics[width=\cmsFigWidth]{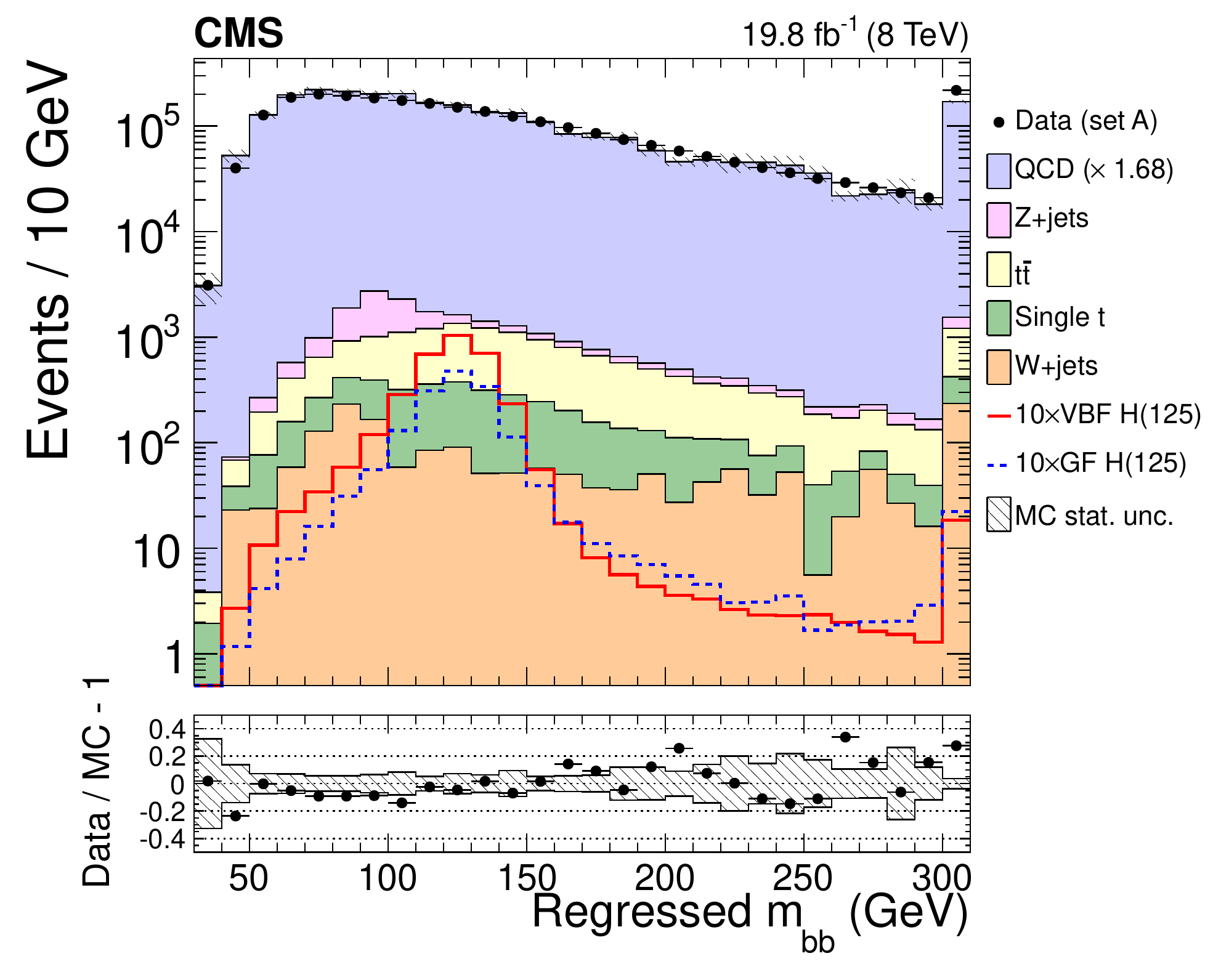}
    \caption{Distribution in invariant mass of the two \PQb-jet candidates, after the jet \pt regression, for the events of set A.
Data are shown by the points, while the simulated backgrounds are stacked. The LO QCD cross section is multiplied by a factor $1.68$ so that the total number of background events equals the number of events in the data, while the VBF and GF Higgs boson signal cross sections are multiplied by a factor 10 for better visibility. The last bin is the sum of all the events beyond the range of the $x$ axis (overflow). The panel at the bottom shows the fractional difference between the data and the background simulation, with the shaded band representing the statistical uncertainties in the MC samples.}
    \label{fig:mbbData}
\end{figure}

The additional energy correction of $\PQb$ jets leads to an improvement of the jet \pt resolution, which in turn improves the dijet invariant mass resolution by approximately 17\% in the phase space of the offline event selections. Figure~\ref{fig:mbbSignal} shows the reconstructed dijet invariant mass of the \PQb-jet candidates ($m_{\PQb\PQb}$) before and after the regression for simulated events passing set A selections. The measured distribution of the regressed $m_{\PQb\PQb}$ in set A is shown in Fig.~\ref{fig:mbbData}.

The validation of the regression technique in data is done with samples of
$\cPZ\to\ell\ell$ events with one or two \cPqb-tagged jets. When the jets are corrected by
  the regression procedure, the \pt\ balance distribution, between the
  $\cPZ$ boson, reconstructed from the leptons, and the \cPqb-tagged jet or dijet
  system is improved to be better centered at zero and narrower than
  when the regression correction is not applied.   In both
  cases the distributions for data and the simulated samples are
  in good agreement after the regression correction is applied.

\subsection{Discrimination between quark- and gluon-originated jets}\label{sec:qgl}

To further identify whether the jet pair with the smallest \PQb-tag values among the four leading jets is likely to originate from hadronization of a light (\PQu,\PQd,\PQs-type) quark, as expected for signal VBF jets, or from gluons, as is more probable for jets produced in QCD processes, a quark-gluon discriminant~\cite{Chatrchyan:2013jya,FSQ-12-035,CMS-PAS-JME-13-002} is applied to the VBF candidate jets.

The discriminant exploits differences in the showering and fragmentation of gluons and quarks,
making use of the following internal jet composition observables based on the PF jet constituents:
(i) the major root-mean square (RMS) of the distribution of jet constituents
in the $\eta$-$\phi$ plane~\cite{CMS-PAS-JME-13-002},
(ii) the minor RMS of the distribution of jet constituents
in the $\eta$-$\phi$ plane~\cite{CMS-PAS-JME-13-002},
(iii) the jet asymmetry pull~\cite{Gallicchio:2010sw},
(iv) the jet particle multiplicity, and
(v) the maximum energy fraction carried by a jet constituent.
The pull and RMS variables are calculated by weighting each jet constituent
by its \pt squared~\cite{CMS-PAS-JME-13-002}.

\begin{figure}[hbt]
  \centering
    \includegraphics[width=\cmsFigWidth]{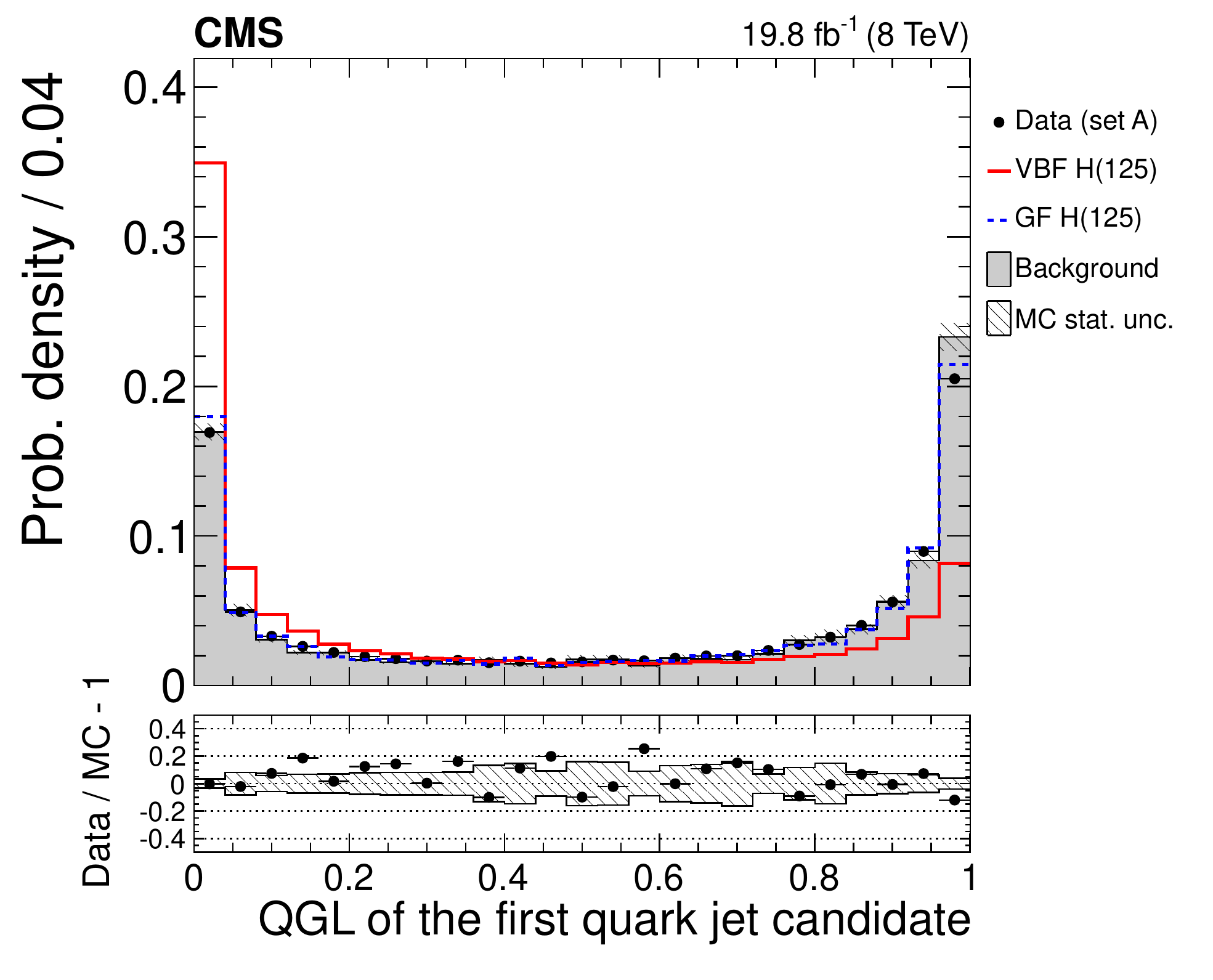}
    \caption{Normalized distribution in quark-gluon likelihood discriminant of the first light-jet candidate. Quark jets are expected to have low likelihood values (closer to 0), while gluon jets are expected to have higher ones (closer to 1). The selection corresponds to set A, data are shown by the points, and the sum of all simulated backgrounds is shown by the filled histogram. The VBF Higgs boson signal is displayed by a solid line, and the GF Higgs boson signal is shown by a dashed line. The panel at the bottom shows the fractional difference between the data and the background simulation, with the shaded band representing the statistical uncertainties in the MC samples.}
    \label{fig:qgl}

\end{figure}

The five variables above are used as inputs to a likelihood estimated with gluon and quark jets from simulated QCD dijet events using the TMVA package. To improve the separation power, all variables are corrected for pileup effects as a function of the \FASTJET $\rho$ density. Figure~\ref{fig:qgl} shows the normalized distribution of the quark-gluon likelihood (QGL)~\cite{CMS-PAS-JME-13-002} for the first VBF-jet candidate (the jet that is ranked third in the \PQb-tag score; see Section~\ref{sec:reco}), for background and signal events. As expected, VBF signal events, dominated by quark jets, have a pronounced peak at likelihood $\sim$0, while the background and GF events are enriched in gluon jets, and
have a very different QGL distribution. The QGL distribution of all four jets is used as input to the signal vs background discriminant (Section~\ref{sec:hmva}).

\subsection{Soft QCD activity}\label{sec:soft}

To measure the additional hadronic activity between the VBF-tagging jets, excluding the more centrally produced Higgs boson decay products, only reconstructed charged tracks are used. This is done to measure the hadronic activity associated with the primary vertex (PV), defined as the reconstructed vertex with the largest sum of squared transverse momenta of tracks used to reconstruct it.

A collection of ``additional tracks'' is assembled using reconstructed tracks that
(i)  satisfy the \textit{high purity} quality requirements defined in
Ref.~\cite{Khachatryan:2010pw} and $\pt>300\MeV$;
(ii) are not associated with any of the four leading PF jets in the event;
(iii) have a minimum longitudinal impact parameter, $\abs{d_z(\mathrm{PV})}$, with respect to the main PV, rather than to other pileup interaction vertices;
(iv) satisfy $\abs{d_z(\mathrm{PV})}<2\unit{mm}$ and $\abs{d_z(\mathrm{PV})}<3\sigma_z(\mathrm{PV})$ with respect to the PV, where $\sigma_z(\mathrm{PV})$ is the uncertainty in $d_z(\mathrm{PV})$; and
(v) are not in the region between the two best \PQb-tagged jets. This is defined as an ellipse in the $\eta$-$\phi$ plane, centered on the midpoint between the two jets, with major axis of length $\Delta R({\PQb\PQb})+1$, where $\Delta R({\PQb\PQb})=\sqrt{(\Delta\eta_{\PQb\PQb})^2+(\Delta\phi_{\PQb\PQb})^2}$, oriented along the direction connecting the two $\PQb$ jets, and with minor axis of length 1.

The additional tracks are then clustered into ``soft TrackJets'' using the anti-\kt clustering algorithm with a distance parameter of 0.5. The use of TrackJets represents a clean and validated method~\cite{CMS-PAS-JME-10-006} to reconstruct the hadronization of partons with very low energies down to a few \GeVns{}~\cite{CMS-PAS-JME-08-001}; an extensive study of the soft TrackJet activity can be found in Refs.~\cite{Chatrchyan:2013jya,FSQ-12-035}.

\begin{figure}[hbt]
  \centering
    \includegraphics[width=\cmsFigWidth]{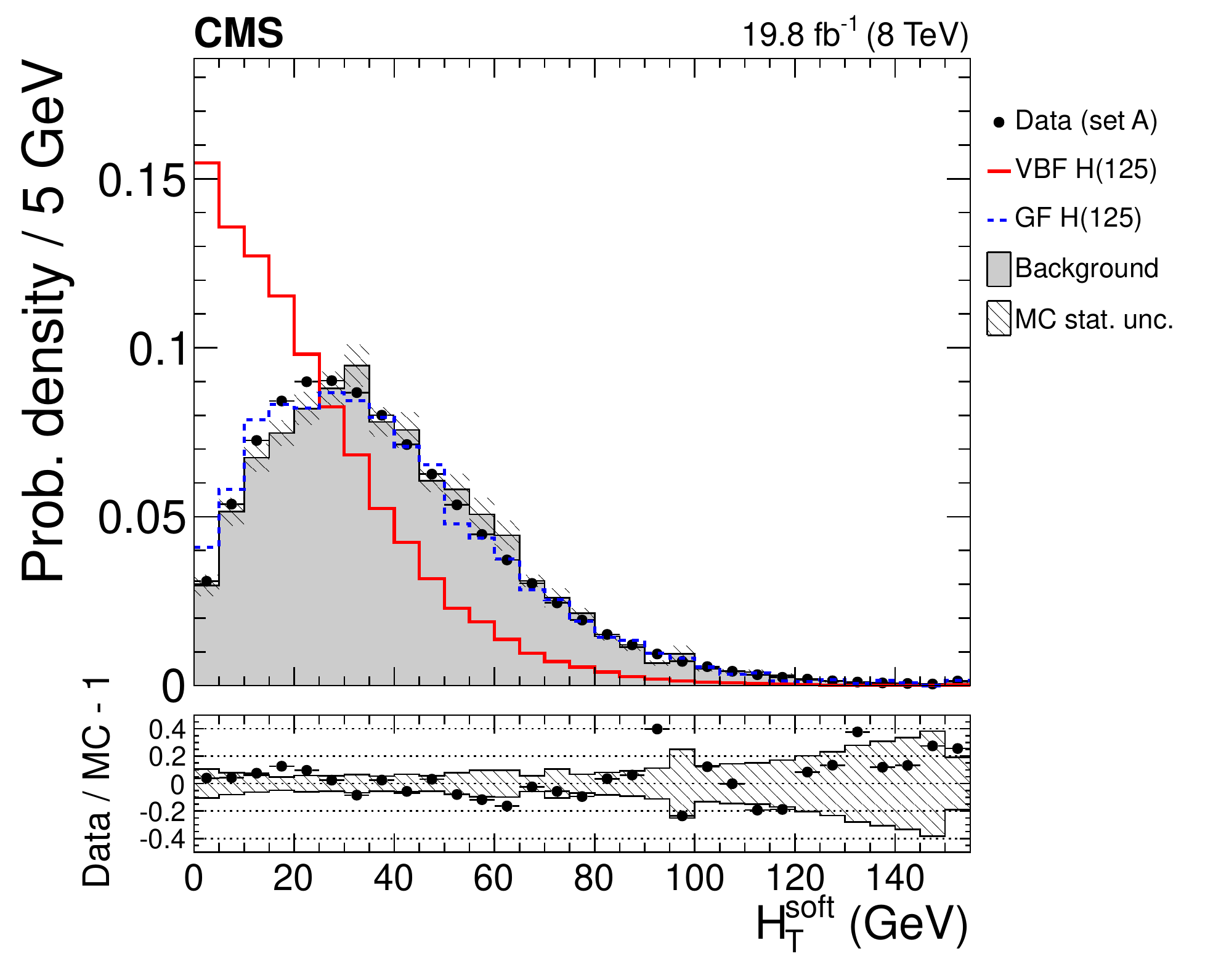}
    \caption{Normalized distribution of the scalar \pt sum of TrackJets that are associated with the soft QCD activity $(\HT^\text{soft})$. The selection corresponds to set A, data are shown by the points, and the sum of all simulated backgrounds is shown by the filled histogram. The VBF Higgs boson signal is displayed by a solid line, and the GF Higgs boson signal is shown by a dashed line. The panel at the bottom shows the fractional difference between the data and the background simulation, with the shaded band representing the statistical uncertainties in the MC samples.}
    \label{fig:softHt}

\end{figure}

The discriminating variable, $\HT^\text{soft}$, that encapsulates the differences between the signal and the QCD background, is the scalar \pt sum of the soft TrackJets with $\pt>1\GeV$, and is shown in Fig.~\ref{fig:softHt}.

\section{Extraction of the \texorpdfstring{$\Z$}{Z} boson signal}\label{sec:z}

The \Z+jets background process, with the $\Z$ boson decaying to a \PQb-quark pair, provides a validation of the analysis strategy used for the Higgs boson search. The extraction of the $\Z$ boson signal demonstrates the ability to observe a relatively wide hadronic resonance on top of a smooth QCD background. Also, if such a signal can be seen, it can serve for \textit{in situ} confirmation of the scale and resolution of the invariant mass of the two $\PQb$ jets. Recently, the observation of a $\Z\to\cPqb\cPaqb$ signal was reported by the ATLAS Collaboration~\cite{ZbbATLAS} in the \Z+1 jet final state, and similar techniques are applied here. The overall strategy has two parts. First, events are selected from set A, with the additional requirement to have at least one CSVM jet. It should be noted that it is important to extract the $\Z$ boson signal in the same four-jet phase space in which the Higgs boson search is performed. Then, a multivariate discriminant is trained to separate the \Z+jets process from the QCD multijet production, using variables that are only weakly correlated to $m_{\PQb\PQb}$. According to the value of the discriminant the events are divided into three categories, ranging from a signal-depleted control category, to a signal-enriched one. Finally, a simultaneous fit of the signal and the QCD background $m_{\PQb\PQb}$ shape is performed in all three categories. The subsequent sections give details of the outlined procedure.

\subsection{\texorpdfstring{$\Z$}{Z} boson signal vs. background discrimination}\label{sec:zmva}

\begin{figure}[hbt]
  \centering
    \includegraphics[width=\cmsFigWidth]{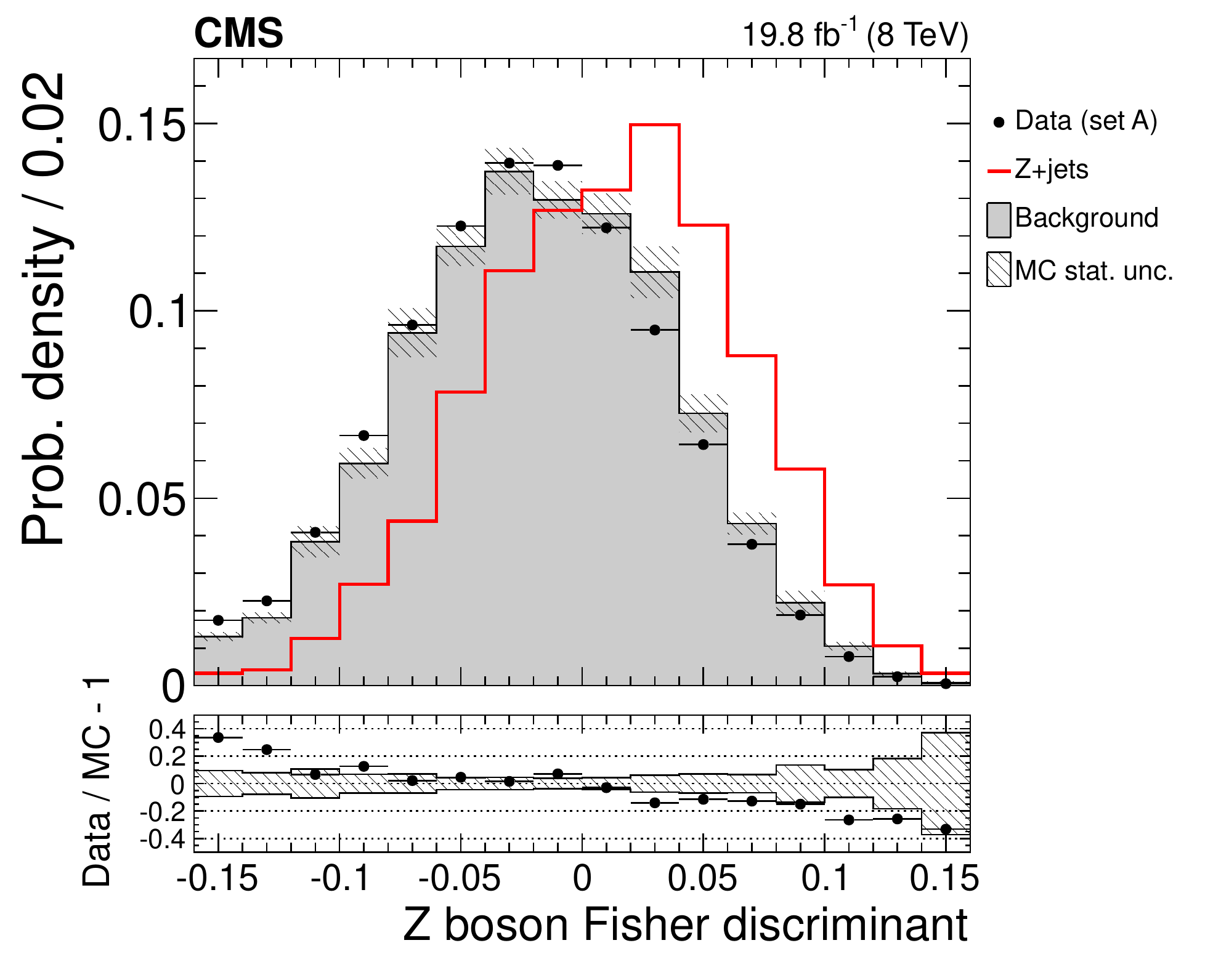}
    \caption{Normalized distribution in $\Z$ boson Fisher discriminant. Data are shown by the points, and the sum of all simulated backgrounds is shown by the filled histogram. The \Z+jets signal is displayed with solid line. The panel at the bottom shows the fractional difference between the data and the background simulation, with the shaded band representing the statistical uncertainties in the MC samples.}
    \label{fig:mvaZ}

\end{figure}

As discussed above, the selection of events is based on set A, with the additional requirement of having at least one CSVM jet; the tightening of the \PQb-tagging condition was found to improve the expected sensitivity.
A Fisher discriminant (FD)~\cite{Hocker:2007ht} is implemented with the TMVA package and trained to discriminate between the \Z+jets signal and the background. For this purpose, seven variables are used:
(i) the absolute $\eta$ difference $\abs{\Delta\eta_{\PQq\PQq}}$ of the VBF jets;
(ii) the absolute $\eta$ of the \PQb-jet system $\abs{\eta_{\PQb\PQb}}$;
(iii) the CSV value of the jet with highest CSV value (with best $\PQb$ tag);
and (iv)-(vii) the QGL values of the four leading jets.
Due to the very small correlations between the variables, the FD performs almost as well as more advanced, nonlinear discriminators.
 Figure~\ref{fig:mvaZ} shows the normalized distribution of the discriminant, where the output of the \Z+jets signal is compared to the background.

\subsection{Fit of the dijet invariant mass spectrum}\label{sec:zfit}

The selected events are divided into three categories, based on the FD output. Table~\ref{tab:Zyields} summarizes the event categories and corresponding yields.

\begin{table}[htbH]
  \centering
    \topcaption{Definition of the event categories for the $\Z$ boson signal extraction and corresponding yields in the $m_{\PQb\PQb}$ interval $[60,170]\GeV$.\label{tab:Zyields}}
    \begin{scotch}{cccc}
            & Category 1 & Category 2 & Category 3 \\
      \hline
            & $\mathrm{FD}<-0.02$ & $-0.02<\mathrm{FD}<0.02$ & $\mathrm{FD}>0.02$ \\
      \hline
      \hline
      Data            	& 659873	& 374797	& 342931  \\
      \Z+jets          	& 1374   	& 1467   	& 2783    \\
      $\ttbar$  	& 2124   	& 1821   	& 2327    \\
      Single \PQt      	& 657    	& 569    	& 812     \\
    \end{scotch}
\end{table}

Besides its discrimination power, the FD has minimal correlation with the invariant mass of the two $\PQb$ jets. This means that the $m_{\PQb\PQb}$ spectrum from QCD processes is independent of the category. In practice, however, there is a small residual dependence, of up to 3\%, which is corrected with a linear transfer function of $m_{\PQb\PQb}$ that is taken from data. The extraction of the $\Z$ boson signal is done with a simultaneous fit in all three categories. Eq.~(\ref{eq:fitZ}) describes the fit model:
\ifthenelse{\boolean{cms@external}}{
\begin{equation}\begin{split}\label{eq:fitZ}
 f_i(m_{\PQb\PQb})=\mu_{\Z}\, N_{i,\Z}\, Z_i(m_{\PQb\PQb};k_\mathrm{JES},k_\mathrm{JER})
 +N_{i,\PQt}\, T_i(m_{\PQb\PQb})\\+N_{i,\mathrm{QCD}}\, K_i(m_{\PQb\PQb})\, B8(m_{\PQb\PQb};\vec{p}),
\end{split}\end{equation}
}{
\begin{equation}\label{eq:fitZ}
 f_i(m_{\PQb\PQb})=\mu_{\Z}\, N_{i,\Z}\, Z_i(m_{\PQb\PQb};k_\mathrm{JES},k_\mathrm{JER})+N_{i,\PQt}\, T_i(m_{\PQb\PQb})+N_{i,\mathrm{QCD}}\, K_i(m_{\PQb\PQb})\, B8(m_{\PQb\PQb};\vec{p}),
\end{equation}
}
where the subscript $i$ denotes the category, $N_{i,\Z}$ is the expected yield for the $\Z$ boson signal; and $\mu_{\Z},\,N_{i,\mathrm{QCD}}$ are free parameters for the signal strength and the QCD event yield. The shape of the top quark background $T_i(m_{\PQb\PQb})$ is taken from the simulation (sum of the $\ttbar$ and single top quark contributions), and the expected yield $N_{i,\PQt}$ is allowed to vary in the fit by 20\%. The \Z+jets signal shape $Z_i(m_{\PQb\PQb};k_\mathrm{JES},k_\mathrm{JER})$ is taken from the simulation and is parametrized as a Crystal Ball function~\cite{Oreglia:1980} (Gaussian core with power-law tail) on top of a combinatorial background modeled by a polynomial. The position and the width of the Gaussian core are allowed to vary by the factors $k_\mathrm{JES}$ and $k_\mathrm{JER}$, respectively, which quantify any mismatch of the jet energy scale and resolution between data and simulation and are constrained by the dedicated validation measurements of the regressed jet energy scale and resolution.

\begin{figure}[hbtp]
  \centering
    \includegraphics[width=0.45\textwidth]{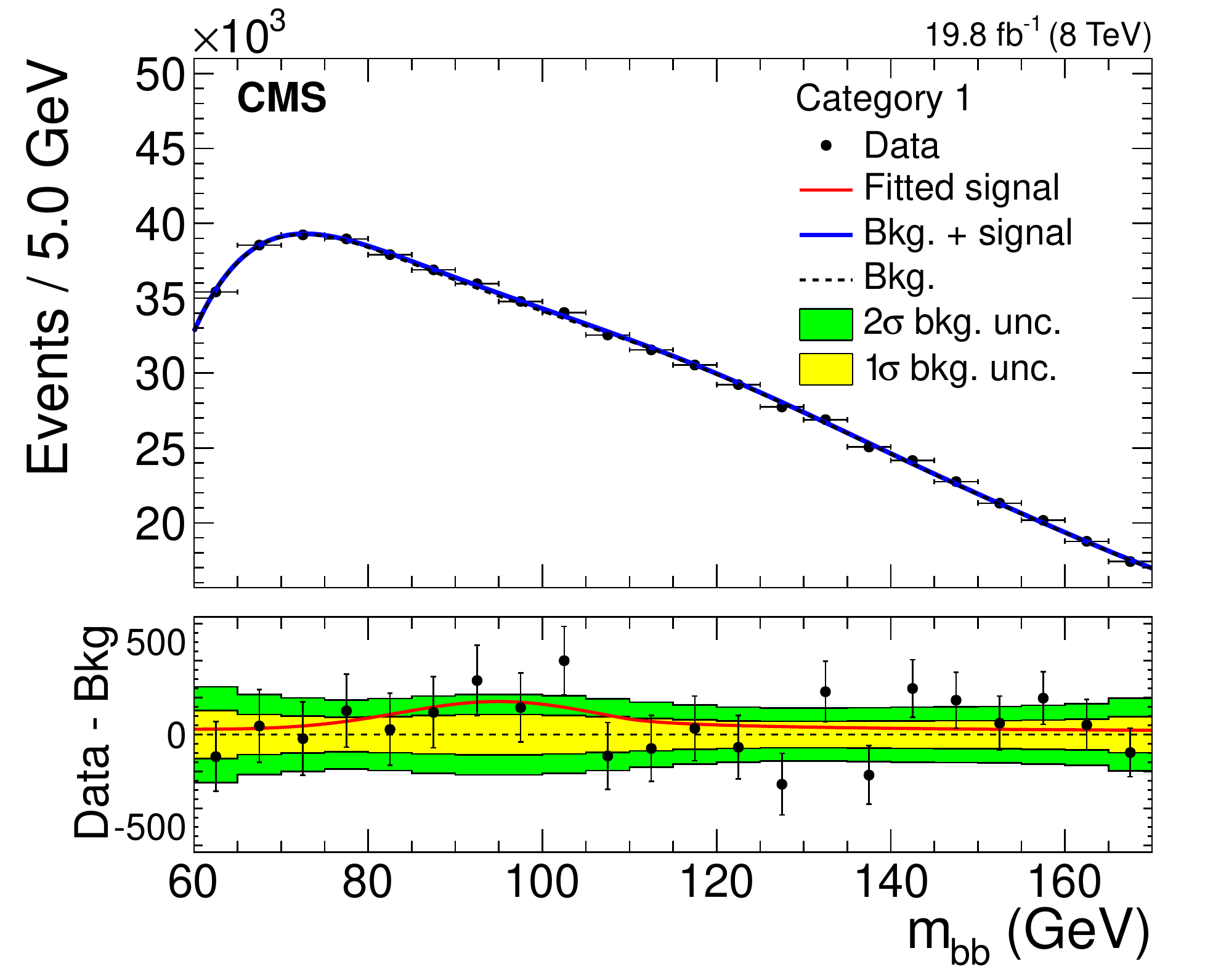}
    \includegraphics[width=0.45\textwidth]{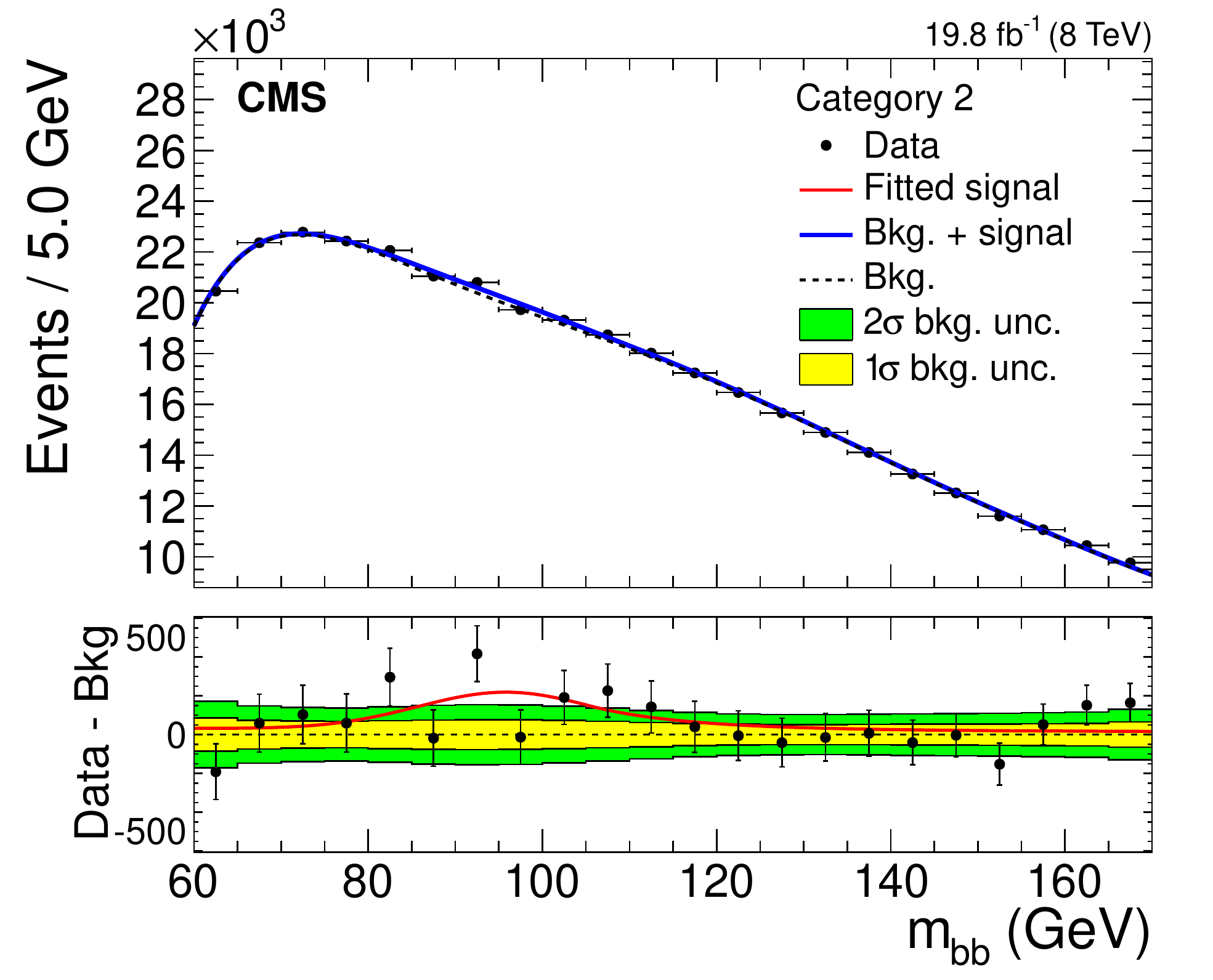}
    \includegraphics[width=0.45\textwidth]{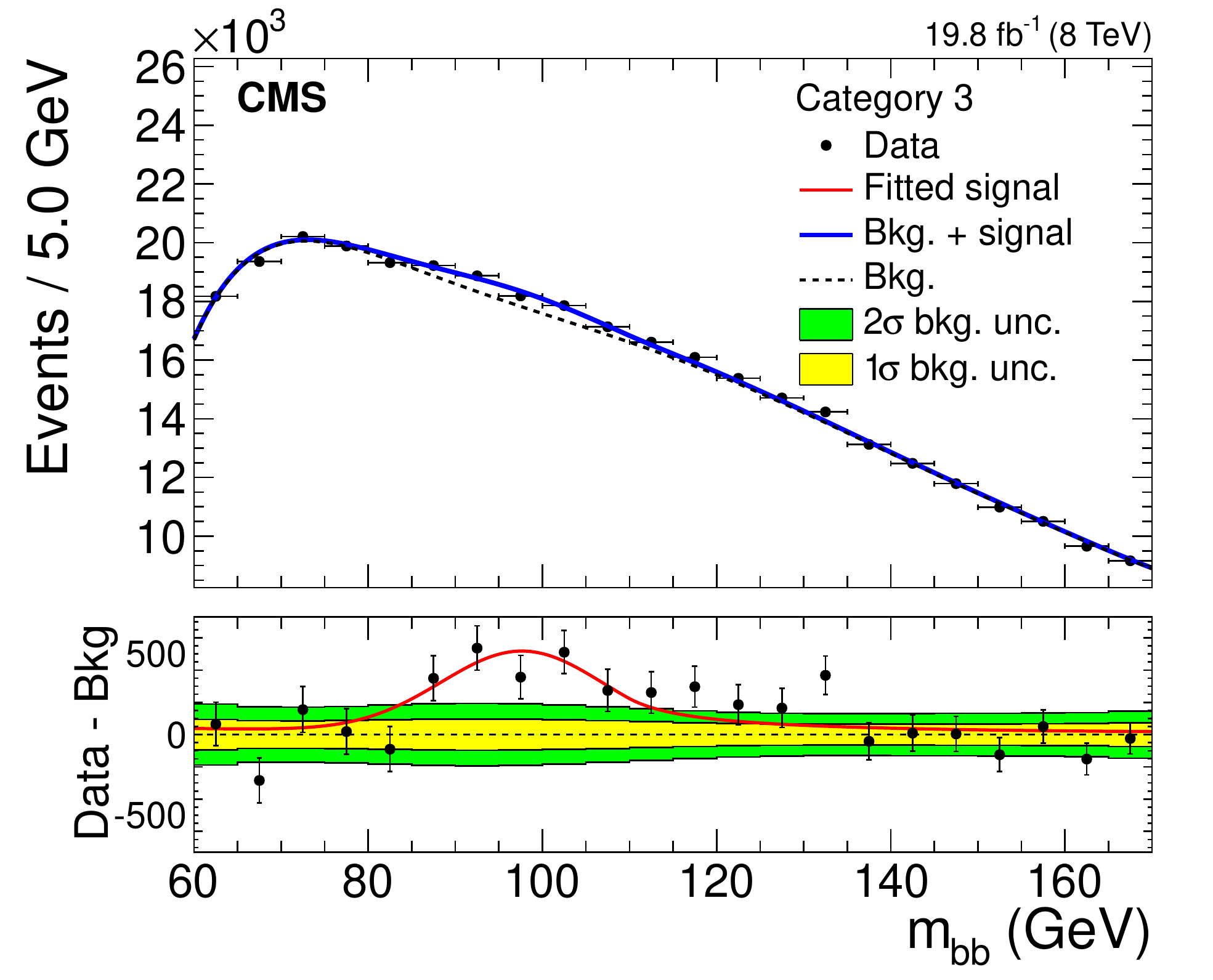}
    \caption{Invariant mass distribution of the two \PQb-jet candidates for the $\Z$ boson signal in the three event categories that are based on the $\Z$ boson Fisher discriminant output, starting from the most backgroundlike (\cmsLeft) and ending at the most signal-like (bottom). Data are shown by the points. The solid line is the sum of the postfit background and signal shapes, while the dashed line is the background component alone. The bottom panel shows the background-subtracted distribution, overlaid with the fitted signal, and with the 1$\sigma$ and 2$\sigma$ background uncertainty bands. The measured (simulated) parameters of the Gaussian core of the signal shape in category 3 are 97.7\,(96.6)\GeV and 9.3\,(9.1)\GeV for the mean and the sigma, respectively.}
    \label{fig:fitZ}
\end{figure}

Finally, the QCD background shape in each category is described by a common, eighth-order polynomial $B8(m_{\PQb\PQb};\vec{p})$, whose parameters $\vec{p}$ are determined by the fit, and a multiplicative linear transfer function $K_i(m_{\PQb\PQb})$ that accounts for the small background shape differences between the categories. The choice of the polynomial is based on an extensive bias study described in Section~\ref{sec:higgs}. Allowing for 20\% uncertainty on the $\Z$ boson signal efficiency, the simultaneous binned maximum-likelihood fit yields a signal strength of $\mu_{\Z}=1.10^{+0.44}_{-0.33}$, which corresponds to an observed (expected) significance of $3.6\,\sigma$ ($3.3\,\sigma$). The fitted values of $k_\mathrm{JES}$ and $k_\mathrm{JER}$ are $1.01\pm 0.02$ and $1.02\pm 0.10$, respectively. Figure~\ref{fig:fitZ} shows the fitted distributions and the background-subtracted ones.

The extraction of the $\Z$ boson signal in this way validates the Higgs boson search method used in this paper
by finding a known dijet resonance in a similar mass range. In addition, the simulated $m_{\PQb\PQb}$ scale and resolution are consistent with the data, based on the best-fit values of the $k_\mathrm{JES}$ and $k_\mathrm{JER}$ nuisance parameters, which serve to constrain the corresponding uncertainties in the Higgs boson signal extraction.

\section{Search for a Higgs boson}\label{sec:higgs}

The search for a Higgs boson follows closely the methodology applied for the extraction of the $\Z$ boson signal (Section~\ref{sec:zfit}). Namely, a multivariate discriminant is employed (Section~\ref{sec:hmva}) to
divide the events into seven categories that are subsequently fit simultaneously with $m_{\PQb\PQb}$ templates (Section~\ref{sec:hfit}).

\subsection{Higgs boson signal vs. background discrimination}\label{sec:hmva}

\begin{figure}[hbt]
  \centering
    \includegraphics[width=0.48\textwidth]{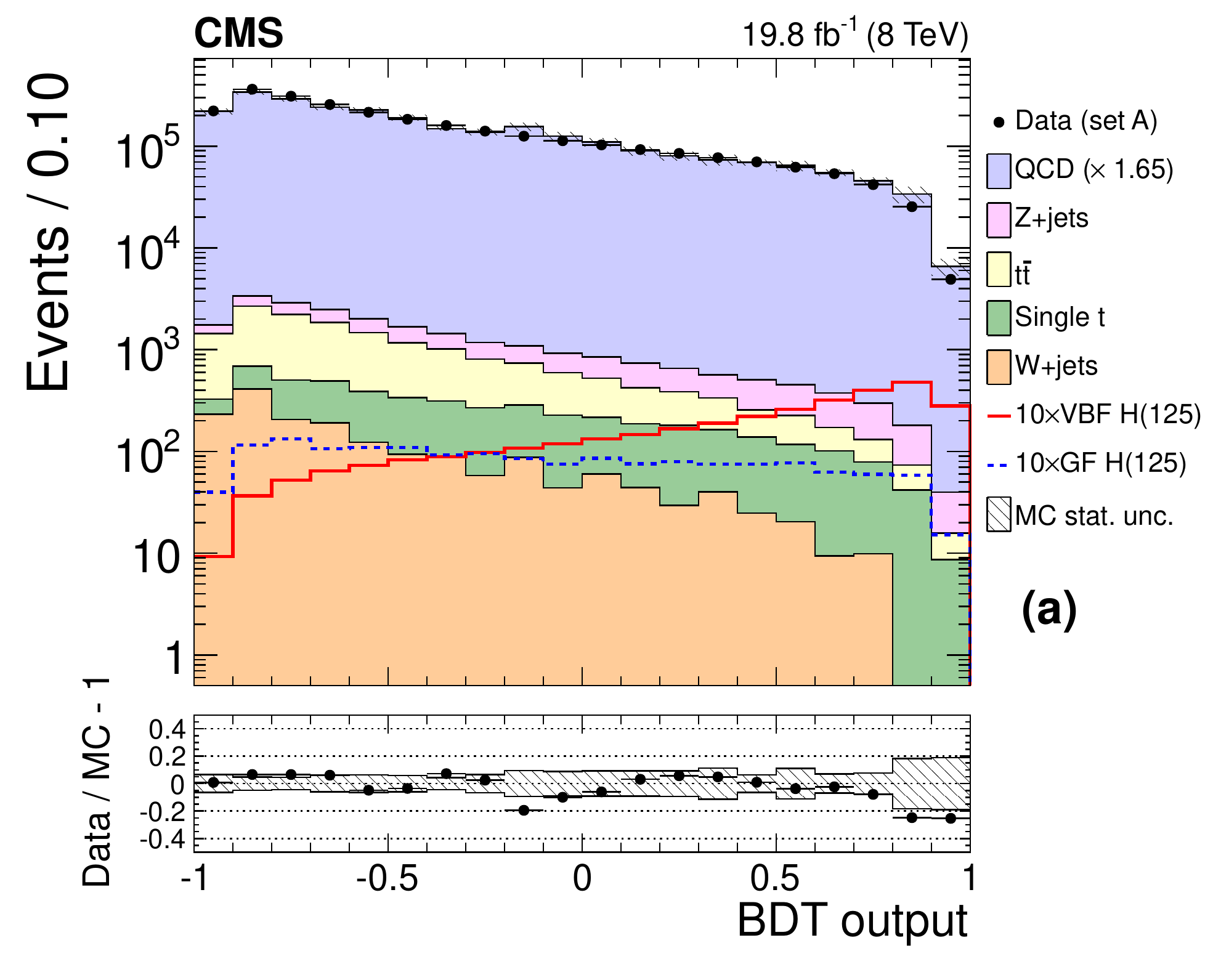}
    \includegraphics[width=0.48\textwidth]{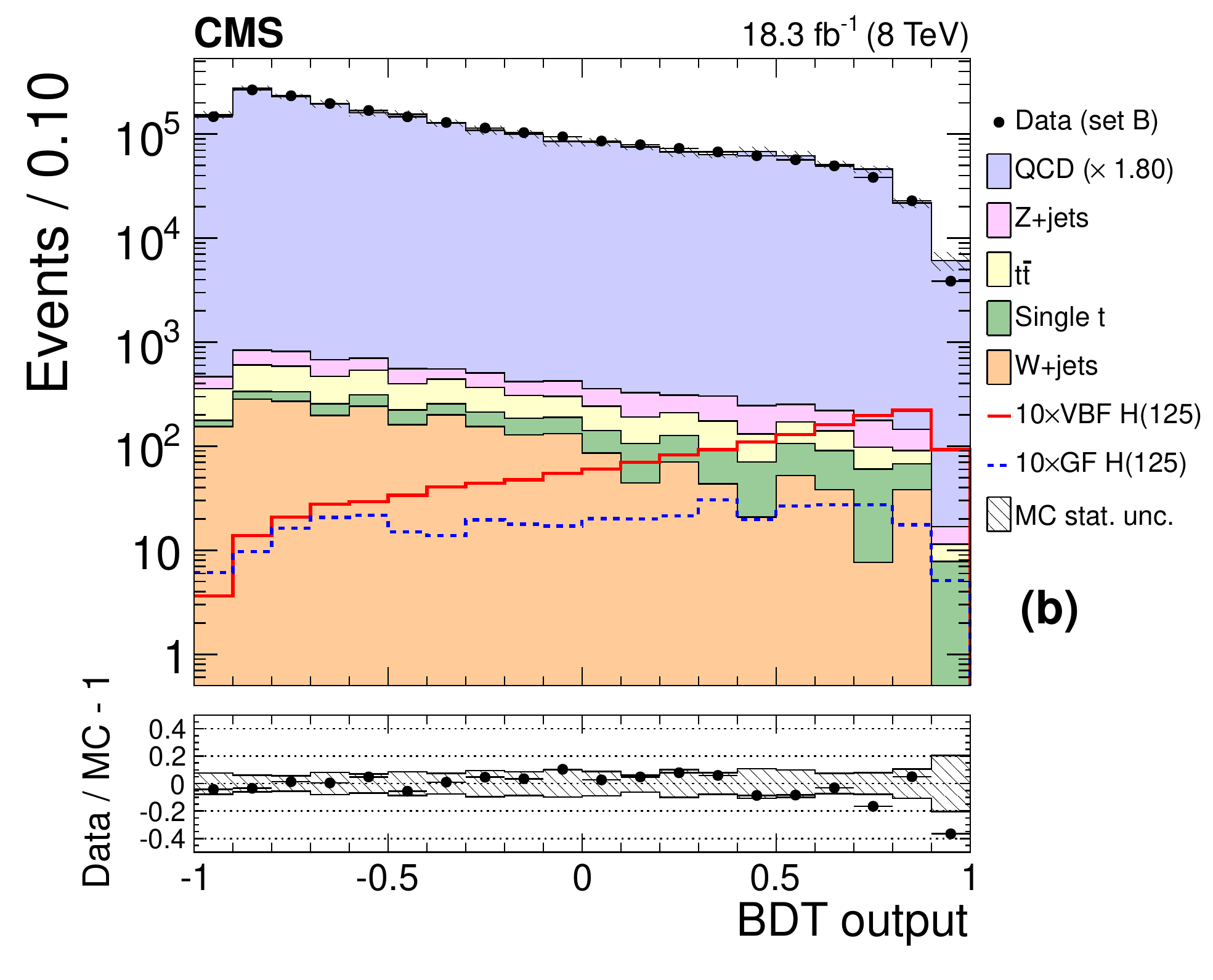}
    \caption{Distribution of the BDT output for the events of set A (a) and set B (b). Data are shown by the points, while the simulated backgrounds are stacked. The LO QCD cross sections are scaled such that the total number of background events equals the number of events in data, while the VBF and GF Higgs boson signal yields are multiplied by a factor of $10$ for better visibility. The panels at the bottom show the fractional difference between the data and the background simulation, with the shaded band representing the statistical uncertainties of the MC samples.}
    \label{fig:mvaHiggs}

\end{figure}

In order to separate the overwhelmingly large QCD background from the Higgs boson signal, all discriminating features have to be used in an optimal way. This is best achieved by using a multivariate discriminant, which in this case is a BDT implemented with the TMVA package. The variables used as an input to the BDT are chosen such that they are very weakly correlated with the dynamics of the $\cPqb\cPaqb$ system, in particular with $m_{\PQb\PQb}$, and are grouped into five distinct groups:
(i) the dynamics of the VBF-jet system, expressed by $\Delta\eta_{\PQq\PQq}$, $\Delta\phi_{\PQq\PQq}$, and $ m_{\PQq\PQq}$;
(ii) the \PQb-jet content of the event, expressed by the CSV output for the two best \PQb-tagged jets;
(iii) the jet flavor of the event QGL for all four jets;
(iv) the soft activity, quantified by the scalar \pt sum $\HT^\text{soft}$ of the additional soft TrackJets with $\pt>1\GeV$,
and the number $N^\text{soft}$ of soft TrackJets with $\pt>2\GeV$;
and (v) the angular dynamics of the production mechanism, expressed by the cosine of the angle between the $\PQq\PQq$ and
$\cPqb\cPaqb$ planes in the center-of-mass frame of the four leading jets $\cos\theta^*_{\PQq\PQq,\PQb\PQb}$.
In practice, two BDTs are trained with the same input variables using the selections corresponding
to the two sets of events.
This distinction is necessary because the properties of the selected events are significantly different between the two selections (set A and set B). Figure~\ref{fig:mvaHiggs} shows the output of the BDT for the two sets of events.

\subsection{Fit of the dijet invariant mass spectrum}\label{sec:hfit}

Taking into account the expected sensitivity of the analysis and the available number of MC events
(necessary to build the various $m_{\PQb\PQb}$ templates), seven categories are defined, according to the BDT output: four for set A and three for set B. The boundaries of the categories and the respective event yields are summarized in Table~\ref{tab:yields}. In an $m_{\PQb\PQb}$ interval of twice the width of the Gaussian core of the signal distribution ($m_{\PH}=125\GeV$), the signal-over-background ratio reaches 1.7\% in the most sensitive category (category 4). It should be noted that both the VBF and GF contributions are added to the Higgs boson signal, with the fraction of the latter ranging from $\sim$50\% in category 1 to $\sim$7\% in category 4.

\begin{table*}[tbH]
\centering
\caption{Definition of the event categories and corresponding yields in the $m_{\PQb\PQb}$ interval $[80,200]\GeV$, for the data and the MC expectation. The BDT output boundary values refer to the distributions shown in Fig.~\ref{fig:mvaHiggs}.\label{tab:yields}}
\resizebox{\textwidth}{!}{
\begin{scotch}{c|cccc|ccc}
\multirow{3}{*}{BDT boundary values} & \multicolumn{4}{c|}{set A} & \multicolumn{3}{c}{set B} \\
\cline{2-8}
                  & Cat. 1 & Cat. 2 & Cat. 3 & Cat. 4 & Cat. 5 & Cat. 6 & Cat. 7 \\
\cline{2-8}
 & $-0.6$ -- 0.0 & 0.0 -- 0.7 & 0.7 -- 0.84 & 0.84 -- 1.0 & $-0.1$ -- 0.4 & 0.4 -- 0.8 & 0.8 -- 1.0 \\
\hline
\hline
Data              & 546121 & 321039 & 32740 & 10874 & 203865 & 108279 & 15151   \\
\hline
\Z+jets            & 2038   & 1584   & 198   & 71    & 435    & 280    & 45      \\
\PW+jets            & 282    & 135    & 4     & $<$1 & 225    & 92     & 17      \\
$\ttbar$     & 2818   & 839    & 45    & 14    &  342   & 169    & 21      \\
Single \PQt	  &  960   & 633    & 64    & 25    & 194    & 159    & 30       \\
\hline
VBF $m_{\PH} (125)$   &  53    & 140    & 58    & 57    &  33    &  57    & 31    \\
GF $m_{\PH} (125)$    &  53    & 51     & 8     &  5    & 9     &   10    & 2  \\
\end{scotch}
}
\end{table*}

The analysis relies on the assumption that the QCD $m_{\PQb\PQb}$ spectrum shape is the same in all BDT categories of the same set of events. In reality, a small correction is needed to account for residual differences between the $m_{\PQb\PQb}$ spectrum in category 1 vs. categories 2,3 and 4, and in category 5 vs. categories 6 and 7. The correction factor (transfer function) is a linear function of $m_{\PQb\PQb}$ in set A and a quadratic one in set B (because a stronger dependence is observed in set B between $m_{\PQb\PQb}$ and the multivariate discriminant). With the introduction of the transfer functions, the fit model for the Higgs boson signal is given by Eq.~(\ref{eq:fitH}):
\ifthenelse{\boolean{cms@external}}{
\begin{equation}\label{eq:fitH}
\begin{split}
 f_i(m_{\PQb\PQb})=&\mu_{\PH}\, N_{i,\PH}\, H_i(m_{\PQb\PQb};k_\mathrm{JES},k_\mathrm{JER})\\
 &+N_{i,\Z}\, Z_i(m_{\PQb\PQb};k_\mathrm{JES},k_\mathrm{JER})\\
 &+N_{i,\PQt}\, T_i(m_{\PQb\PQb};k_\mathrm{JES},k_\mathrm{JER})\\
 &+N_{i,\mathrm{QCD}}\, K_i(m_{\PQb\PQb})\, B(m_{\PQb\PQb};\vec{p}_\text{set}),
\end{split}
\end{equation}
}{
\begin{equation}\label{eq:fitH}
\begin{split}
 f_i(m_{\PQb\PQb})=\mu_{\PH}\, N_{i,\PH}\, H_i(m_{\PQb\PQb};k_\mathrm{JES},k_\mathrm{JER})+N_{i,\Z}\, Z_i(m_{\PQb\PQb};k_\mathrm{JES},k_\mathrm{JER})\\
 +N_{i,\PQt}\, T_i(m_{\PQb\PQb};k_\mathrm{JES},k_\mathrm{JER})+N_{i,\mathrm{QCD}}\, K_i(m_{\PQb\PQb})\, B(m_{\PQb\PQb};\vec{p}_\text{set}),
\end{split}
\end{equation}
}
where the subscript $i$ denotes the category and $\mu_{\PH},\,N_{i,\mathrm{QCD}}$ are free parameters for the signal strength and the QCD event yield. $N_{i,\PH}$, $N_{i,\Z}$, and $N_{i,\PQt}$ are the expected yields for the Higgs boson signal, the \Z+jets, and the top quark background respectively.
The shape of the top quark background $T_i(m_{\PQb\PQb};k_\mathrm{JES},k_\mathrm{JER})$ is taken from the simulation
(sum of the $\ttbar$ and single top quark contributions) and is described by a broad Gaussian. The Z/W+jets background $Z_i(m_{\PQb\PQb};k_\mathrm{JES},k_\mathrm{JER})$ and the Higgs boson signal $H_i(m_{\PQb\PQb};k_\mathrm{JES},k_\mathrm{JER})$ shapes are taken from the simulation and are parametrized as a Crystal Ball function (Gaussian core with power-law tail) on top of a polynomial. The position and the width of the Gaussian core of the MC templates (signal and background) are allowed to vary by the free factors $k_\mathrm{JES}$ and $k_\mathrm{JER}$, respectively, which quantify any mismatch of the jet energy scale and resolution between data and simulation.
Finally, the QCD shape is described by a polynomial $B(m_{\PQb\PQb};\vec{p}_\text{set})$, common within the categories of each set,
and a multiplicative transfer function $K_i(m_{\PQb\PQb})$ per category, accounting for the shape differences between the categories.
The parameters of the polynomial, $\vec{p}_\text{set}$, and those of the transfer functions, are determined by the fit, which is performed simultaneously in all categories in each set. For set A, the polynomial is of fifth order, while for set B it is of fourth order.

The choice of the QCD background shapes and event category transfer function parametrizations are fully based on data, and have been performed among classes of functions, e.g. polynomials, exponential, power laws and their combinations, with a minimum number of degrees of freedom suited to fit the data in all categories. Each function considered is used to generate different MC pseudo-data sets, and each data set is fitted using the different functional models. A potential bias on the signal estimation is computed  for each pair of possible functions used to generate and fit to the pseudo-data sets.
The background model chosen yields a maximum potential bias on the fitted signal strength of less than six times the statistical uncertainty on the background. Hence the systematic uncertainty associated with the background shape can be neglected.

\section{Systematic uncertainties}\label{sec:unc}
\setlength{\parskip}{6.0pt}

Table~\ref{tab:systematics}  summarizes the sources of uncertainty related to both the background and to the signal processes.
The leading uncertainty comes from the QCD background description: both the parameters of its shape and the overall normalization in each category are allowed to float freely, being determined by the simultaneous fit to the data. The resulting covariance matrix is used to compute the uncertainty. For the smaller background contributions from the \Z/\PW+jets and top quark production, the $m_{\PQb\PQb}$ shapes are taken from the simulation, while their corresponding yields are allowed to float in the fit with a 30\% log-normal constraint centered on the SM expectations.

The experimental uncertainties on the jet energy scale (JES) and jet energy resolution (JER) affect the signal acceptance and the shape of the multivariate discriminant output, and are included as nuisance parameters. The effect of the JES and JER uncertainties on the $m_{\PQb\PQb}$ shape is taken into account in the fit function. By varying the JES and JER by their measured uncertainties~\cite{JES}, the impact of the signal yield per analysis category is estimated. These variations affect the acceptance by up to 10\%, while the peak position of the $m_{\PQb\PQb}$ shape is shifted by 2\%, and the width by 10\%.

Additional uncertainties are assigned to the flavor tagging of the jets. The CSV and QGL discriminant outputs are shifted according to the observed agreement between data and simulation
and the effect on the signal acceptance is estimated to range from 3\% to 10\% for the former, and from 1\% to 3\% for the latter. The impact of the CSV shift is more significant, both because it is used for the event selection, and because the multivariate discriminant depends more strongly on the $\PQb$ tagging of the jets. The shift of QGL only affects the shape of the discriminant, and thus the distribution of signal events in the analysis categories.

The trigger uncertainty is estimated by propagating the uncertainty in the data vs. MC simulation scale factor for the efficiency.
This is achieved by convolving the two-dimensional efficiency scale factor with the signal distribution. As a result, the uncertainty in the signal yield ranges from 1\% to 6\% for the VBF process, and from 5\% to 20\% for the GF.

Theoretical uncertainties affect the signal simulation. First, the uncertainty due to PDFs and strong coupling constant $\alpha_S$ variation is computed to be 2.8\% (VBF) and 7.5\% (GF)~\cite{Heinemeyer:2013tqa}. A residual uncertainty from these sources is estimated for the particular kinematical phase space of the search: following the PDF4LHC prescription~\cite{PDF4LHC1,PDF4LHC2} the PDF and $\alpha_S$ uncertainty ranges from 2\% to 5\%, while the renormalization and factorization scale variations in the signal simulation induce an uncertainty of 1\% to 5\% in the analysis categories, on top of a global cross section uncertainty of 0.2\% (VBF) and 7.7--8.1\% (GF). Finally, the variation of the UE and parton-shower (PS) model (using \PYTHIA8.1~\cite{Sjostrand:2007gs} instead of the default \PYTHIA6) affects the signal acceptance by 2\% to 7\% (VBF) and by 10\% to 45\% (GF).

Lastly, an uncertainty of 2.6\% is assigned to the total integrated luminosity measurement~\cite{CMS-PAS-LUM-13-001}.\vskip -0.4em

\setlength{\parskip}{6.8pt}
\begin{table}[h]
\centering
\topcaption{Sources of systematic uncertainty and their impact on the shape and normalization of the background and signal processes. \label{tab:systematics}}
\begin{scotch}{p{0.01cm}lrr}

 \multicolumn{4}{l}{Background uncertainties} \\
 \hline
 &QCD shape parameters   		& \multicolumn{2}{c}{determined by the fit} \\
 &QCD bkg. normalization       		& \multicolumn{2}{c}{determined by the fit} \\
 &Top quark bkg. normalization 		& \multicolumn{2}{c}{30\%} \\
 &Z/W+jets bkg. normalization 	        & \multicolumn{2}{c}{30\%} \\
 \hline\hline
 \multicolumn{2}{l}{Uncertainties affecting the signal} & \multicolumn{1}{c}{VBF signal} & \multicolumn{1}{c}{GF signal} \\
 \hline
 &JES (signal shape)           		& \multicolumn{2}{c}{2\%} \\
 &JER (signal shape)          		& \multicolumn{2}{c}{10\%} \\
 &Integrated luminosity                	& \multicolumn{2}{c}{2.6\%} \\
 &Branching fraction ($\PH\to\cPqb\cPaqb$) 	& \multicolumn{2}{c}{2.4\%--4.3\%}\\
 \hline
 &JES (acceptance)              	& 6\%--10\% 	& 4\%--12\%          \\
 &JER (acceptance)            		& 1\%--4\%   	&   1\%--9\%        \\
 &\PQb-jet tagging                		&  3\%--9\%  	& 3\%--10\%         \\
 &Quark/gluon-jet tagging 		& 1\%--3\% 	& 1\%--3\%          \\
 &Trigger                  		& 1\%--6\% 	& 5\%--20\%          \\
 \hline\hline
 \multicolumn{2}{l}{Theory uncertainties} & \multicolumn{1}{c}{VBF signal} & \multicolumn{1}{c}{GF signal} \\
 \hline
 &UE \& PS                      	& 2\%--7\% 	& 10\%--45\%          \\
 &Scale variation (global) 		& 0.2\% 	& 7.7\%--8.1\% \\
 &Scale variation (categories) 		& 1\%--5\% 	&1\%--5\% \\
 &PDF (global)				& 2.8\% 	& 7.5\%   \\
 &PDF (categories)			& 1.5\%--3\% 	& 3.5\%--5\%   \\
\end{scotch}
\end{table}

\section{Results}\label{sec:res}

\begin{figure*}[hbtp]
  \centering
    \includegraphics[width=0.48\textwidth]{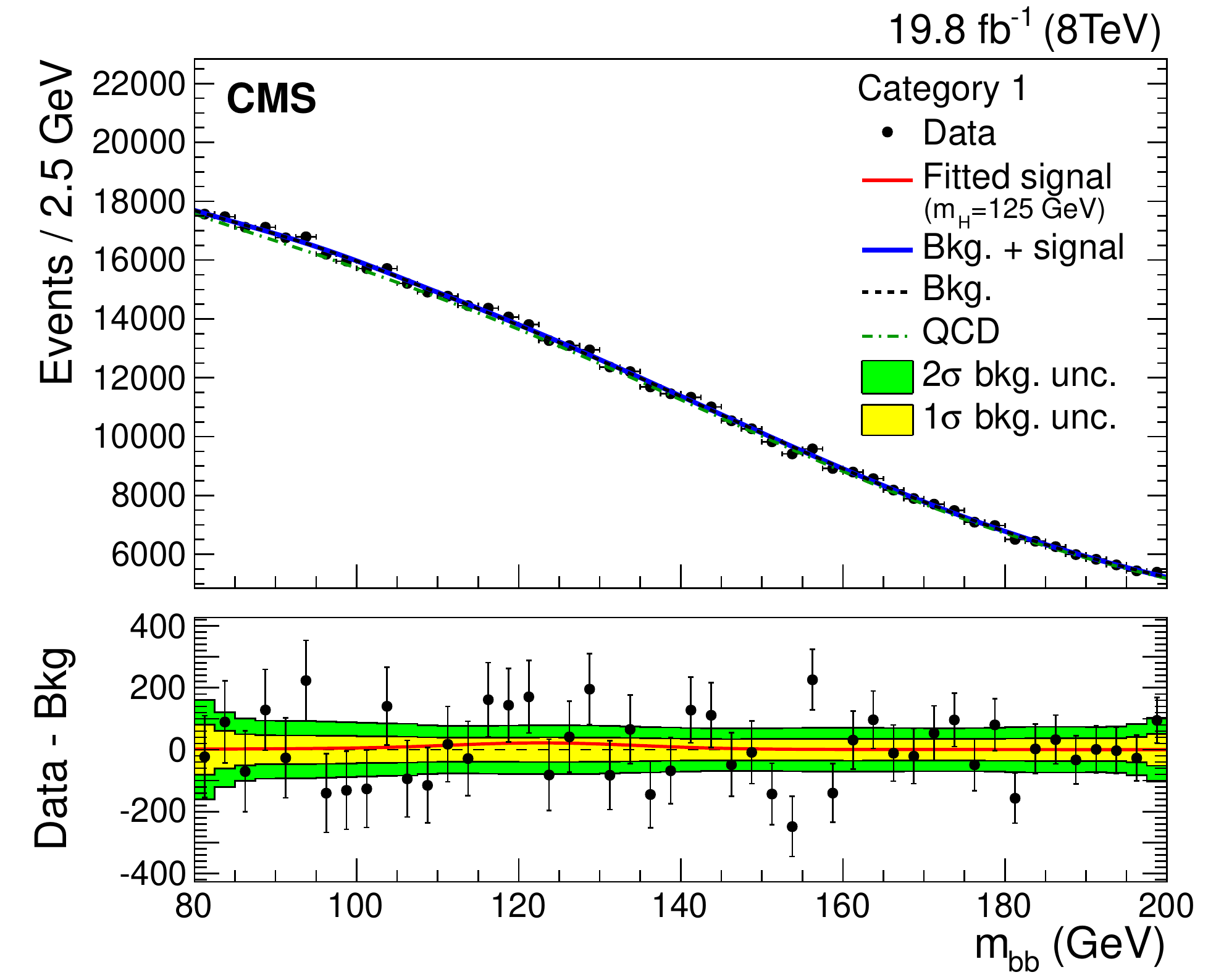}
    \includegraphics[width=0.48\textwidth]{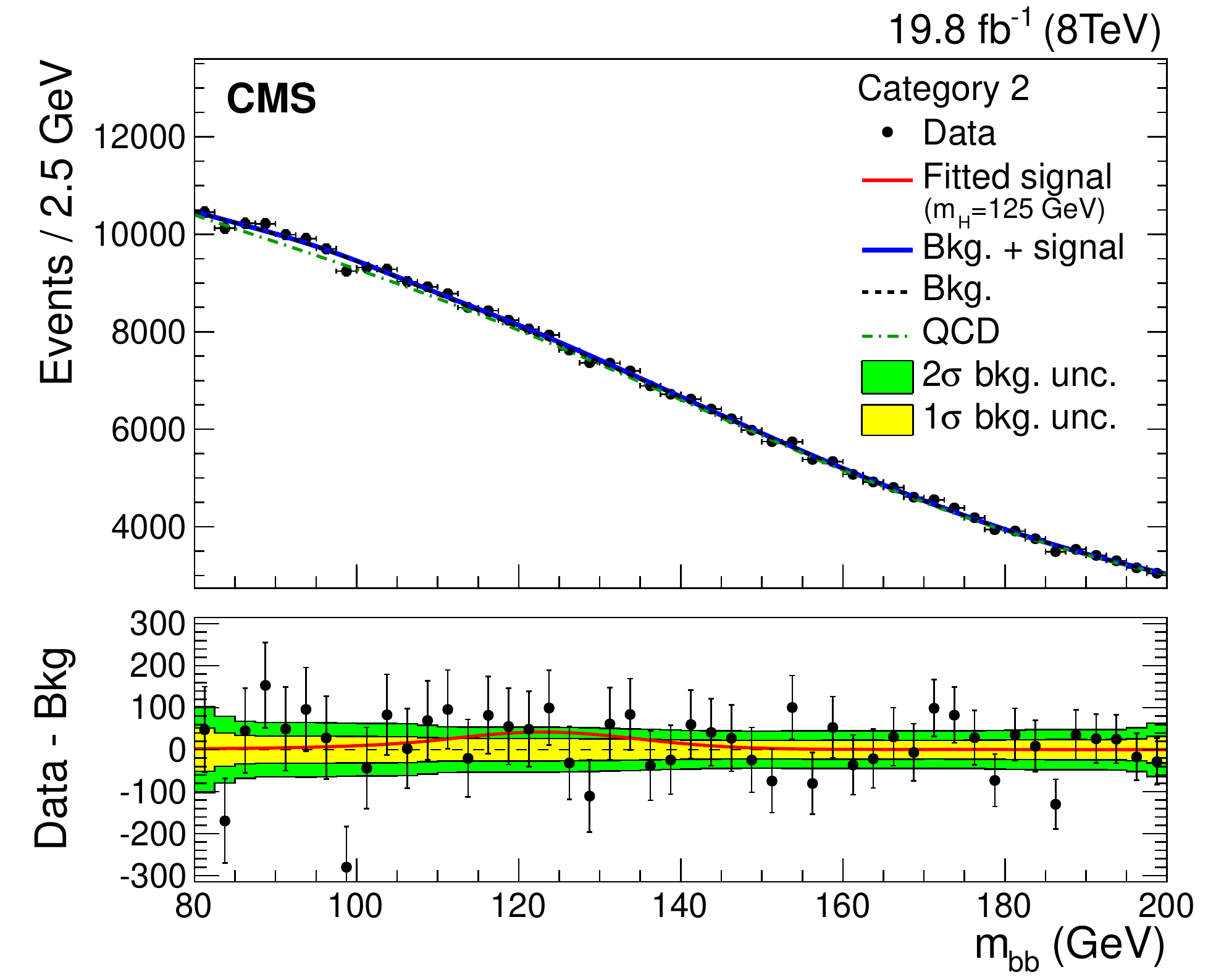}
    \includegraphics[width=0.48\textwidth]{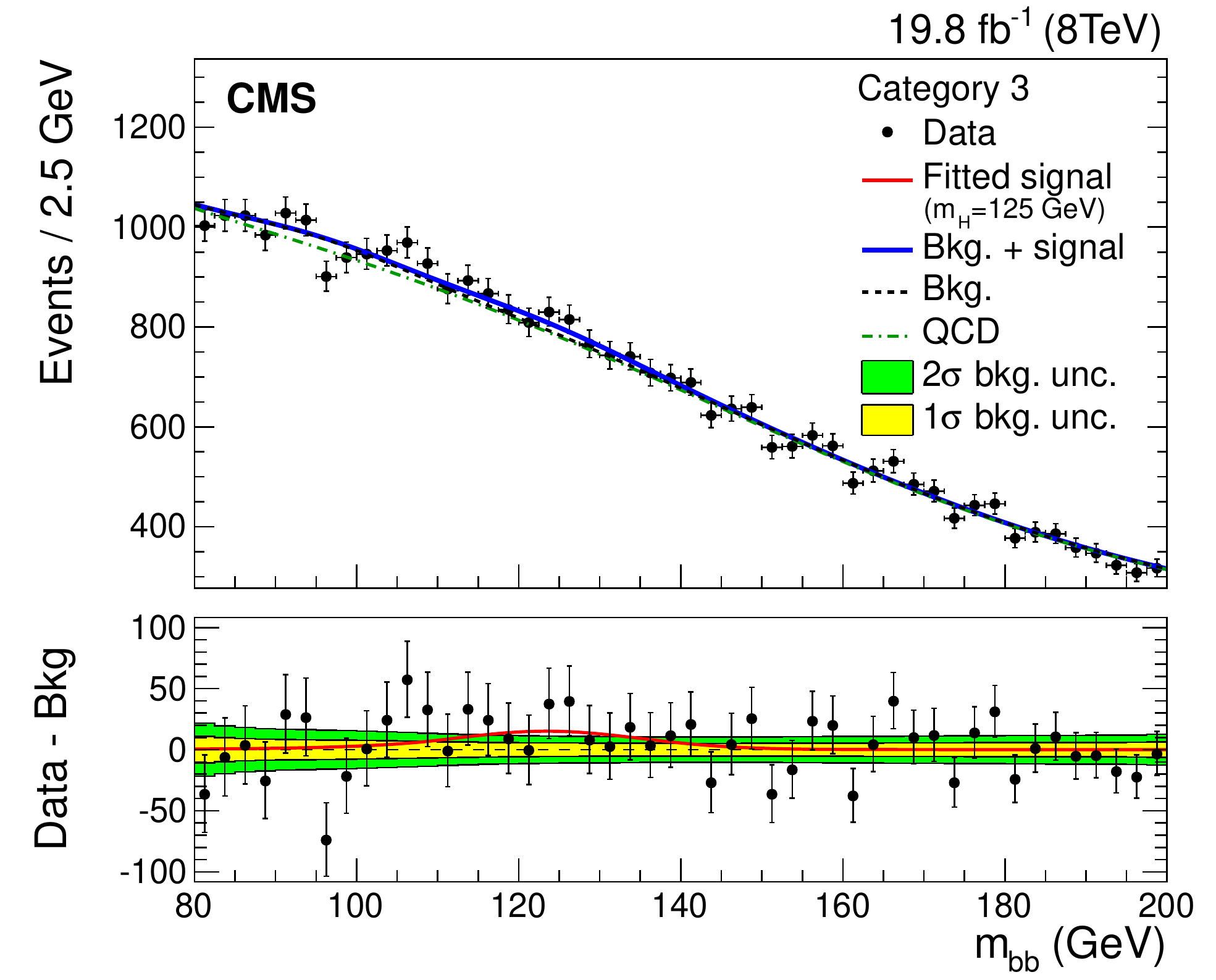}
    \includegraphics[width=0.48\textwidth]{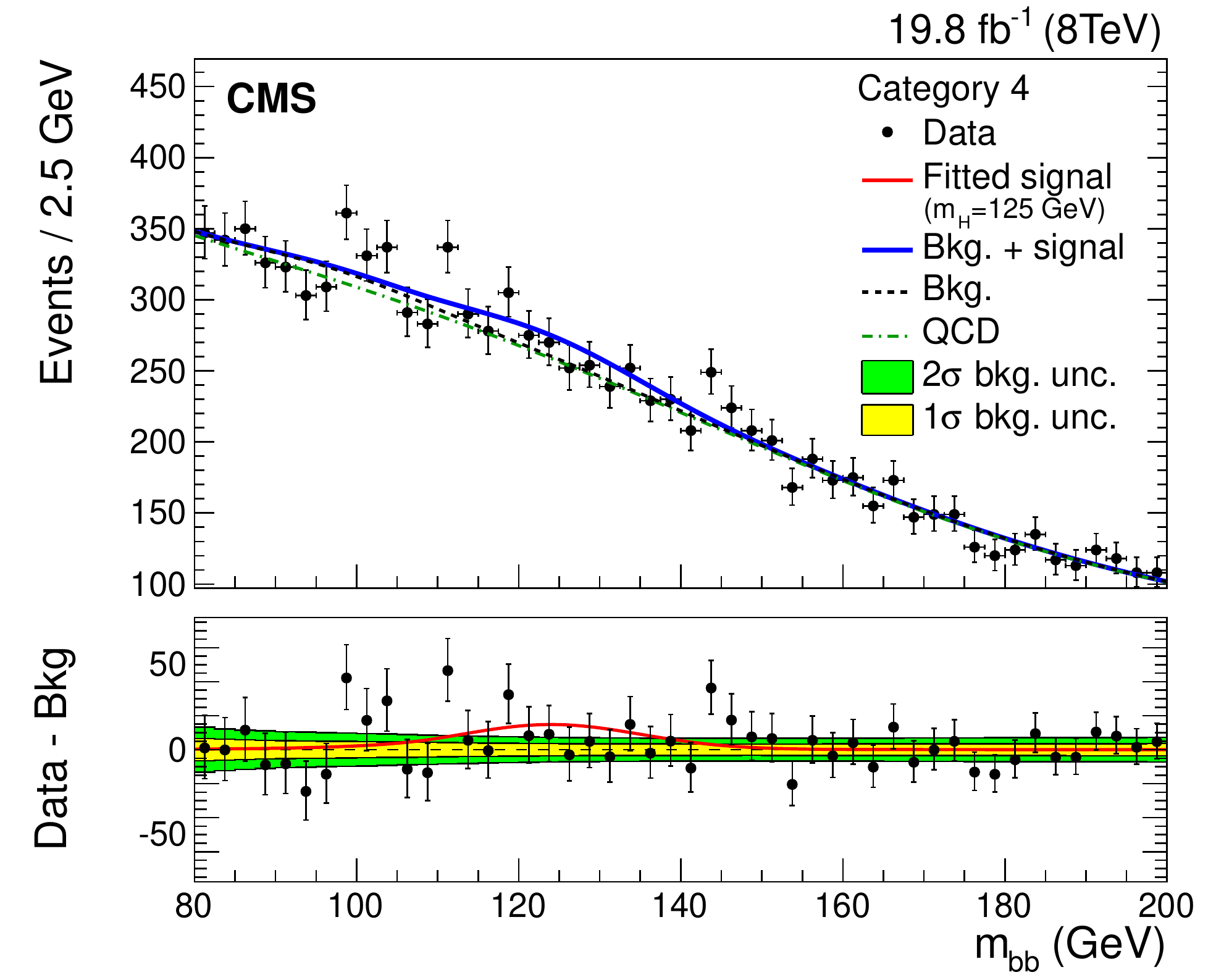}
    \caption{Fit of the invariant mass of the two \PQb-jet candidates for the Higgs boson signal ($m_{\PH}=125\GeV$) in the four event categories of set A. Data are shown by the points. The solid line is the sum of the postfit background and signal shapes, the dashed line is the background component, and the dashed-dotted line is the QCD component alone. The bottom panel shows the background-subtracted distribution, overlaid with the fitted signal, and with the 1$\sigma$ and 2$\sigma$ background uncertainty bands.}
    \label{fig:fitHiggsNOM}
\end{figure*}

\begin{figure}[hbt]
  \centering
    \includegraphics[width=0.45\textwidth]{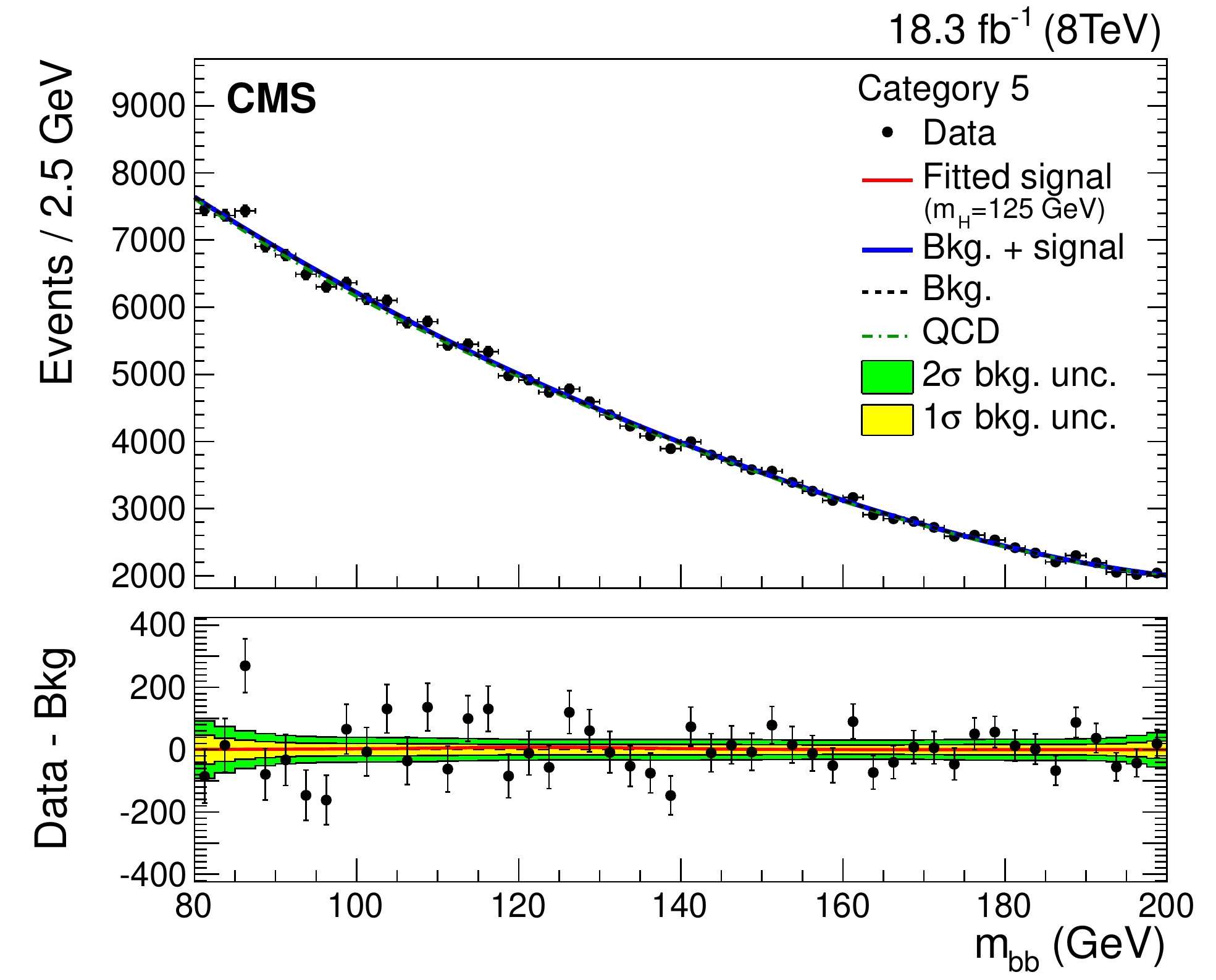}
    \includegraphics[width=0.45\textwidth]{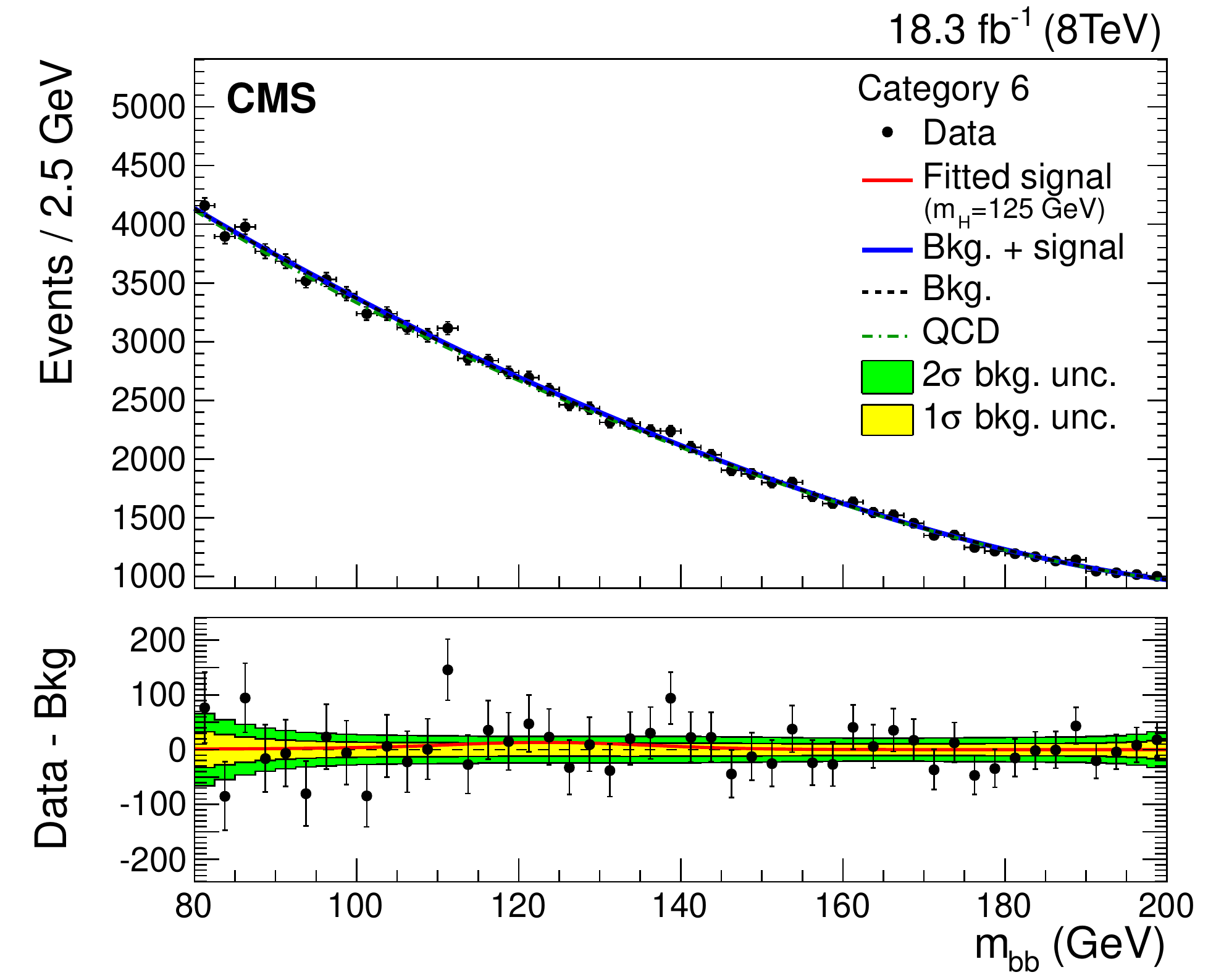}
    \includegraphics[width=0.45\textwidth]{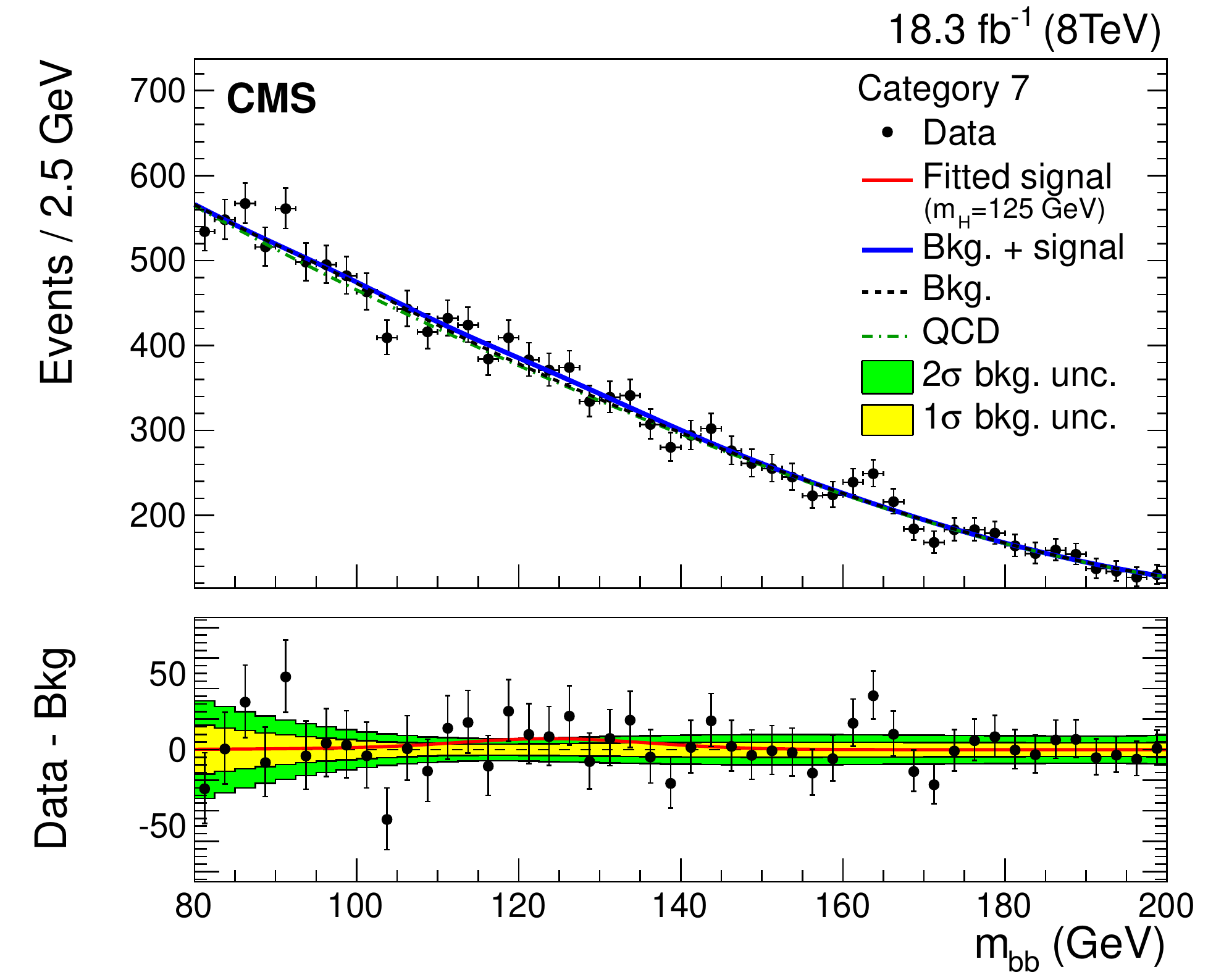}
    \caption{Fit of the invariant mass of the two \PQb-jet candidates for the Higgs boson signal ($m_{\PH}=125\GeV$) in the three event categories of set B. Data are shown with markers. The solid line is the sum of the postfit background and signal shapes, the dashed line is the background component, and the dashed-dotted line is the QCD component alone. The bottom panel shows the background-subtracted distribution, overlaid with the fitted signal, and with the 1$\sigma$ and 2$\sigma$ background uncertainty bands.}
    \label{fig:fitHiggsVBF}
\end{figure}

\begin{figure}[hbtp]
  \centering
    \includegraphics[width=\cmsFigWidth]{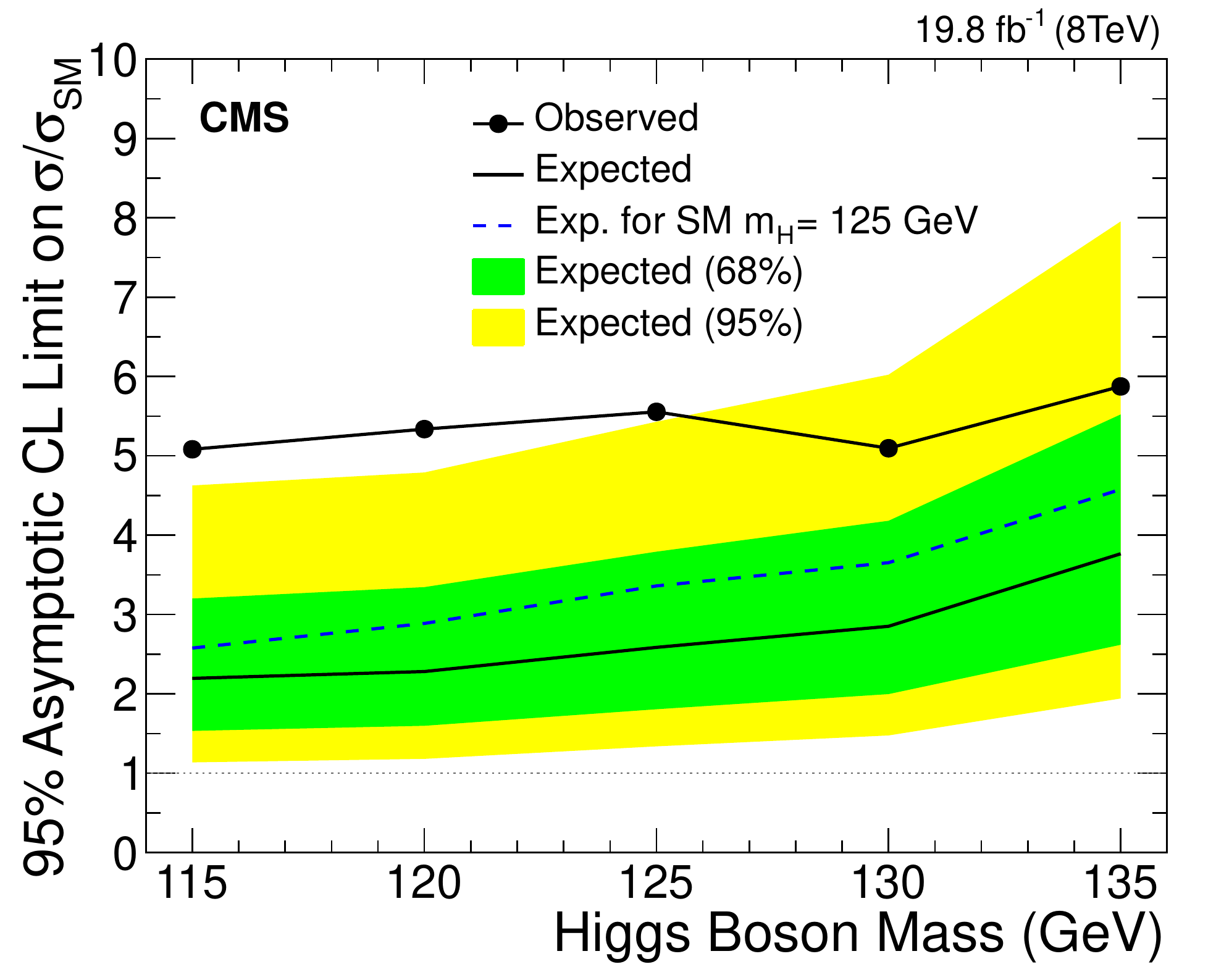}
    \caption{Expected and observed 95\% confidence level limits on the signal cross section in units of the SM expected cross section,
as a function of the Higgs boson mass, including all event categories. The limits expected in the presence of a SM Higgs boson with a mass of 125\GeV are indicated by the dotted curve.}
    \label{fig:limits}
\end{figure}

The $m_{\PQb\PQb}$ distributions in data, for all categories, are fitted simultaneously with the parametric functions described in Section~\ref{sec:hfit} under two different hypotheses: background only and background plus a Higgs boson signal. The fit is a binned likelihood fit incorporating the systematic uncertainties discussed in Section~\ref{sec:unc} as nuisance parameters. Due to the smallness of the GF contribution in the most signal-sensitive categories we do not attempt to fit independently the VBF and the GF signal strengths. The fits in sets A and B are shown in Figs.~\ref{fig:fitHiggsNOM} and~\ref{fig:fitHiggsVBF}, respectively. The limits on the signal strength are computed with the asymptotic \CLs method~\cite{bib:CLsJunk,bib:CLsRead,bib:asCLs}. Figure~\ref{fig:limits} shows the observed (expected) 95\% confidence level (\CL) limit on the total VBF plus GF signal strength, as a function of the Higgs boson mass, which ranges from $5.0$ ($2.2$) at $m_{\PH}=115\GeV$ to 5.8 (3.7) at $m_{\PH}=135\GeV$, together with the expected limits in the presence of a SM Higgs boson with a mass of 125\GeV. For the 125\GeV Higgs boson signal the observed (expected) significance is 2.2 (0.8) standard deviations, and the fitted signal strength is $\mu=\sigma/\sigma_\mathrm{SM}=2.8^{+1.6}_{-1.4}$. The measured signal strength is compatible with the SM Higgs boson prediction $\mu=1$ at the 8\% level.

\section{Combination with other CMS Higgs boson to \PQb-quarks searches}\label{sec:comb}
The CMS experiment has also performed searches for the Higgs boson decaying to bottom quarks, where the Higgs boson is produced in association with a vector boson~\cite{CMS-VHbb} (VH), or with a top quark pair~\cite{CMS-ttHbb,CMS-ttH} ($\ttbar\PH$). The VH results have been recently updated and combined with $\ttbar\PH$~\cite{Khachatryan:2014jba}.
Here we combine those results with the ones from the VBF production search described in this paper.
Event selection overlaps between different analyses have been checked and are either empty by construction or have negligible effects on the combination. The combination methodology is based on the likelihood ratio test statistics employed in Section~\ref{sec:res}, and takes into account correlations among sources of systematic uncertainty. Care is taken to understand the behavior of the parameters that are correlated between analyses, in terms of the fitted parameter values and uncertainties.

\begin{table*}[htb]
\centering
\topcaption{Observed and expected 95\% \CL limits, best fit values on the signal strength parameter $\mu = \sigma/\sigma_\mathrm{SM}$ and signal significances for $m_{\PH}= 125\GeV$, for each $\PH\to\cPqb\cPaqb$ channel and their combination.\label{tab:comb}}
\begin{scotch}{cccccc}
$\PH\to\cPqb\cPaqb$& Best fit (68\% \CL) &  \multicolumn{2}{c}{Upper limits (95\% \CL) } &  \multicolumn{2}{c}{Signal significance} \\
 Channel & Observed &  Observed &  Expected &  Observed &  Expected \\
 \hline
 \hline
VH 		& $0.89\pm 0.43$ 		& 1.68 	& 0.85 	& 2.08 & 2.52 \\
$\ttbar\PH$ & $0.7\pm 1.8$ 			& 4.1 	& 3.5	& 0.37 & 0.58 \\
VBF 		& $2.8^{+1.6}_{-1.4}$ 		& 5.5 	& 2.5	& 2.20 & 0.83 \\
 \hline
Combined	&  $1.03^{+0.44}_{-0.42}$ 	&  1.77 &  0.78 & 2.56 & 2.70 \\
\end{scotch}
\end{table*}

Table~\ref{tab:comb} lists the 95\% \CL expected and observed upper limits and the best-fit signal strength values from the individual channels and from the combined fit. For $m_{\PH}= 125\GeV$ the combination yields an $\PH\to\cPqb\cPaqb$ signal strength $\mu=1.03^{+0.44}_{-0.42}$ with a significance of 2.6 standard deviations.

\section{Summary}\label{sec:sum}

A search has been carried out for the SM Higgs boson produced in vector boson fusion and decaying to $\cPqb\cPaqb$ with two data samples of $\Pp\Pp$ collisions at $\sqrt{s}=8\TeV$ collected with the CMS detector at the LHC corresponding to integrated luminosities of 19.8\fbinv and 18.3\fbinv. Upper limits, at the 95\% confidence level, on the production cross section times the $\PH\to\cPqb\cPaqb$ branching fraction, relative to expectations for a SM Higgs boson, are extracted for a Higgs boson in the mass range 115--135\GeV. In this range, the expected upper limits in the absence of a signal vary between a factor of 2.2 to 3.7 of the SM prediction, while the observed upper limits vary from 5.0 to 5.8. For a Higgs boson mass of 125\GeV, the observed and expected significance is, respectively, 2.2 and 0.8 standard deviations, and the fitted signal strength is $\mu=\sigma/\sigma_\mathrm{SM}=2.8^{+1.6}_{-1.4}$. This is the first search of this kind, and the only search for the SM Higgs boson in all-jet final states, at the LHC.

The combination of the results obtained in this search with other CMS $\PH\to\cPqb\cPaqb$ searches in the VH and $\ttbar\PH$ production modes, yields a $\PH\to\cPqb\cPaqb$ signal strength $\mu=1.03^{+0.44}_{-0.42}$ with a signal significance of 2.6 standard deviations for $m_{\PH}= 125\GeV$ that is consistent with the SM.

\clearpage
\begin{acknowledgments}
We congratulate our colleagues in the CERN accelerator departments for the excellent performance of the LHC and thank the technical and administrative staffs at CERN and at other CMS institutes for their contributions to the success of the CMS effort. In addition, we gratefully acknowledge the computing centers and personnel of the Worldwide LHC Computing Grid for delivering so effectively the computing infrastructure essential to our analyses. Finally, we acknowledge the enduring support for the construction and operation of the LHC and the CMS detector provided by the following funding agencies: BMWFW and FWF (Austria); FNRS and FWO (Belgium); CNPq, CAPES, FAPERJ, and FAPESP (Brazil); MES (Bulgaria); CERN; CAS, MoST, and NSFC (China); COLCIENCIAS (Colombia); MSES and CSF (Croatia); RPF (Cyprus); MoER, ERC IUT and ERDF (Estonia); Academy of Finland, MEC, and HIP (Finland); CEA and CNRS/IN2P3 (France); BMBF, DFG, and HGF (Germany); GSRT (Greece); OTKA and NIH (Hungary); DAE and DST (India); IPM (Iran); SFI (Ireland); INFN (Italy); MSIP and NRF (Republic of Korea); LAS (Lithuania); MOE and UM (Malaysia); CINVESTAV, CONACYT, SEP, and UASLP-FAI (Mexico); MBIE (New Zealand); PAEC (Pakistan); MSHE and NSC (Poland); FCT (Portugal); JINR (Dubna); MON, RosAtom, RAS and RFBR (Russia); MESTD (Serbia); SEIDI and CPAN (Spain); Swiss Funding Agencies (Switzerland); MST (Taipei); ThEPCenter, IPST, STAR and NSTDA (Thailand); TUBITAK and TAEK (Turkey); NASU and SFFR (Ukraine); STFC (United Kingdom); DOE and NSF (USA).

Individuals have received support from the Marie-Curie program and the European Research Council and EPLANET (European Union); the Leventis Foundation; the A. P. Sloan Foundation; the Alexander von Humboldt Foundation; the Belgian Federal Science Policy Office; the Fonds pour la Formation \`a la Recherche dans l'Industrie et dans l'Agriculture (FRIA-Belgium); the Agentschap voor Innovatie door Wetenschap en Technologie (IWT-Belgium); the Ministry of Education, Youth and Sports (MEYS) of the Czech Republic; the Council of Science and Industrial Research, India; the HOMING PLUS program of the Foundation for Polish Science, cofinanced from the European Union, Regional Development Fund; the Compagnia di San Paolo (Torino); the Consorzio per la Fisica (Trieste); MIUR project 20108T4XTM (Italy); the Thalis and Aristeia programs cofinanced by EU-ESF and the Greek NSRF; and the National Priorities Research Program by the Qatar National Research Fund.
\end{acknowledgments}

\bibliography{auto_generated}

\cleardoublepage \appendix\section{The CMS Collaboration \label{app:collab}}\begin{sloppypar}\hyphenpenalty=5000\widowpenalty=500\clubpenalty=5000\textbf{Yerevan Physics Institute,  Yerevan,  Armenia}\\*[0pt]
V.~Khachatryan, A.M.~Sirunyan, A.~Tumasyan
\vskip\cmsinstskip
\textbf{Institut f\"{u}r Hochenergiephysik der OeAW,  Wien,  Austria}\\*[0pt]
W.~Adam, E.~Asilar, T.~Bergauer, J.~Brandstetter, E.~Brondolin, M.~Dragicevic, J.~Er\"{o}, M.~Flechl, M.~Friedl, R.~Fr\"{u}hwirth\cmsAuthorMark{1}, V.M.~Ghete, C.~Hartl, N.~H\"{o}rmann, J.~Hrubec, M.~Jeitler\cmsAuthorMark{1}, V.~Kn\"{u}nz, A.~K\"{o}nig, M.~Krammer\cmsAuthorMark{1}, I.~Kr\"{a}tschmer, D.~Liko, T.~Matsushita, I.~Mikulec, D.~Rabady\cmsAuthorMark{2}, B.~Rahbaran, H.~Rohringer, J.~Schieck\cmsAuthorMark{1}, R.~Sch\"{o}fbeck, J.~Strauss, W.~Treberer-Treberspurg, W.~Waltenberger, C.-E.~Wulz\cmsAuthorMark{1}
% ...
\vskip\cmsinstskip
% ...
1:~~Also at Vienna University of Technology, Vienna, Austria\\
2:~~Also at CERN, European Organization for Nuclear Research, Geneva, Switzerland\\
% ...

\end{sloppypar}
\end{document}